\documentclass[11pt]{article}
\usepackage{comment}
\usepackage[margin=2.5cm]{geometry}
\usepackage[toc,page]{appendix}
\usepackage{amsmath,amsfonts,breqn,comment,cancel,mathtools,amssymb,extarrows,mathtools,graphicx,subfigure,setspace,bbold,physics,bm}
\usepackage{xr-hyper} 
\usepackage{cite}
\usepackage{slashed}
\usepackage[dvipsnames]{xcolor}
\usepackage{hyperref}
\hypersetup{colorlinks=true, linkcolor=blue, citecolor=magenta, linktoc=page}
\usepackage{url}
\usepackage[normalem]{ulem}
\usepackage{tikz}
\usepackage{color}
\usepackage{arydshln,leftidx,mathtools}
\usetikzlibrary{decorations.pathreplacing}
\numberwithin{equation}{section}
\newcommand{\be}{\begin{equation}}
	\newcommand{\bea}{\begin{eqnarray}}
		\newcommand{\eea}{\end{eqnarray}}
	\newcommand{\ba}{\begin{align}}
		\newcommand{\ea}{\end{align}}
	\newcommand{\ee}{\end{equation}}

\usepackage{chngcntr}
\begin{document}
	\onehalfspacing
\begin{titlepage}
	\thispagestyle{empty}

%
	
	\vspace{.4cm}
\begin{center}{\Large \textbf{
Symmetry resolved entanglement in Lifshitz field theories
}}\end{center}

\vspace*{15mm}
		\vspace*{1mm}

\begin{center}
{M. Reza Mohammadi Mozaffar$^a$ and  Ali Mollabashi$^b$ 
}
\end{center}
\vspace*{1cm}

\begin{center}
{\it $^a$ Department of Physics, University of Guilan, P.O. Box 41335-1914, Rasht, Iran\\
  $^b$ School of Quantum Physics and Matter, Institute for Research in Fundamental Sciences (IPM), P.O.Box 19395-5531, Tehran, Iran
		}
		
		\vspace*{0.5cm}
		{E-mails: {\tt mmohammadi@guilan.ac.ir, mollabashi@ipm.ir}}%
\end{center}

\begin{abstract}
We investigate symmetry-resolved entanglement in non-relativistic quantum field theories, including complex Lifshitz scalar chains and Lifshitz fermionic models. Using charged moments and the correlator method, we compute symmetry-resolved Rényi and von Neumann entropies and analyze their dependence on subsystem size, charge, mass, and the dynamical exponent $z$. Our results reveal distinct features of non-relativistic entanglement. In Lifshitz scalar theories, approximate equipartition among charge sectors emerges in the large-$z$ regime, with configurational entropy dominating, whereas Lifshitz fermionic models exhibit genuine equipartition only in the relativistic limit, with fluctuation entropy prevailing. These findings highlight a rich interplay between conserved charges, subsystem size, mass, and dynamical scaling, and provide a framework to explore operationally accessible entanglement in non-relativistic systems. Our study offers insights relevant to experimental platforms such as cold atom setups and mesoscopic systems, where particle-number-resolved measurements can probe symmetry-resolved entanglement.
\end{abstract}

\end{titlepage}
\newpage

\tableofcontents
\noindent
\hrulefill


\section{Introduction}\label{intro}

Entanglement is a central concept in quantum physics, providing a quantitative measure of correlations between subsystems and serving as a key tool in quantum information theory, many-body physics, and quantum field theory \cite{book:2000,Calabrese:2009qy,Laflorencie:2015eck,Nishioka:2018khk,Casini:2022rlv,Rangamani:2016dms}. While conventional entanglement measures characterize the total quantum correlations in a subsystem, they do not distinguish contributions arising from different symmetry sectors. \textit{Symmetry-resolved entanglement measures} address this limitation by decomposing the entanglement according to conserved quantities associated with global symmetries, such as particle number, spin, or charge. The first systematic treatment of entanglement resolution based on symmetry was pioneered in \cite{Laflorencie:2014cem}, laying the groundwork for studying the distribution of quantum information across different charge sectors. This refinement not only deepens our understanding of how quantum correlations are organized but also provides insights into charge fluctuations, transport properties, and the role of symmetries in critical and topological systems \cite{Castro-Alvaredo2024}. Furthermore, symmetry-resolved entanglement in such systems is directly relevant to experimental platforms, including cold atom setups and mesoscopic systems, where conserved particle numbers can be measured and controlled, making these theoretical insights experimentally accessible \cite{Lukin:2019,Neven:2021igr,Rath:2022qif}. See also \cite{Goldstein:2017bua,Xavier:2018kqb,Bonsignori:2019naz,Tan:2019axb,Murciano:2019wdl,Capizzi2020,Murciano:2020lqq,Murciano:2020vgh,Murciano:2021djk,Capizzi:2021zga,Ares:2022hdh,Jones:2022tgp,Ghasemi:2022jxg,Capizzi2022,Capizzi:2022nel,Foligno:2022ltu,DiGiulio:2022jjd,Northe:2023khz,DiGiulio:2023nvz,Murciano:2023ofp,Caputa:2025mii,Northe:2025qcv,Ares:2026vjt} for several developments over the past few years on symmetry-resolved entanglement and information measures in both critical and gapped systems. Let us also note that holographic aspects of symmetry-resolved entanglement have been explored in two-dimensional CFTs, including the role of $U(1)$ Chern--Simons couplings \cite{Zhao:2020qmn} and extensions to excited states and multiple entangling intervals \cite{Weisenberger:2021eby}. Higher spin symmetries have also been considered in this context \cite{Zhao:2022wnp}.

In this work, we aim to compute symmetry-resolved entanglement measures within the framework of \textit{non-relativistic quantum field theories (QFTs)}, focusing on models with Lifshitz scaling symmetry. These theories are distinguished by their unconventional scaling properties: Lifshitz models exhibit anisotropic scaling between space and time characterized by a dynamical critical exponent $z \neq 1$ (see \cite{Alexandre:2011kr,Baiguera2023} for reviews). Quantum entanglement and its various measures have been widely studied in non-relativistic quantum field theories and many-body models both in static and time-dependent setups, revealing behaviors that differ markedly from those in relativistic conformal systems 
\cite{Fradkin:2006mb,Solodukhin:2009sk,Nesterov:2010yi,Keranen:2011xs,Zhou:2016ykv,MohammadiMozaffar:2017nri,He:2017wla,MohammadiMozaffar:2017chk,MohammadiMozaffar:2018vmk,MohammadiMozaffar:2019gpn,Angel-Ramelli:2020wfo,Mollabashi:2020ifv,Mozaffar:2021nex,Boudreault:2021pgj,Mintchev:2022xqh,Mintchev:2022yuo,Berthiere:2023bwn,Basak:2023otu,Vasli:2024mrf,Khoshdooni:2025ddk}. 

Understanding how entanglement distributes across different symmetry sectors in non-relativistic systems provides deeper insight into the structure of quantum correlations, the interplay between conserved charges and non-relativistic symmetries, and the emergence of universal features in entanglement scaling. To address this problem, we consider a quantum system prepared in a pure state described by a density matrix $\rho$. Suppose the theory admits a global internal symmetry generated by a conserved charge operator $Q$. Now if the system is in a representation of the charge, then $[\rho, Q]=0$. Further, for a bipartition of the system into a subsystem $A$ and its complement $\bar{A}$, the reduced density matrix is obtained by tracing out the complement, i.e., $\rho_A = \mathrm{Tr}_{\bar{A}} \rho$. In general, the charge operator splits as $Q=Q_A+Q_B$, with each term arising from the local degrees of freedom of the corresponding subsystem. The crucial property underlying \textit{symmetry resolution} is that the reduced density matrix commutes with the charge operator restricted to subsystem $A$,
\begin{equation}
[\rho_A, Q_A] = 0,
\end{equation}
which ensures that $\rho_A$ is block-diagonal in the eigenbasis of $Q_A$, allowing for a decomposition into charge sectors as follows
\begin{equation}\label{rhoAdecom}
\rho_A=\oplus_q\Pi_q \rho_A=\oplus_q\left(p(q) \rho_A(q)\right),
\end{equation}
where $p(q)=\mathrm{Tr}\!\left(\rho_A \Pi_{q} \right)$ denotes the probability of finding the subsystem in sector $q$. Formally, the decomposition is implemented through the \textit{projector operator} onto the subspace with fixed charge eigenvalue $q$,
\begin{equation}
\Pi_{q} = \int_{-\pi}^{\pi} \frac{d\alpha}{2\pi}\, e^{i\alpha(Q_A-q)}.
\end{equation}
Using this projector, the reduced density matrix in a given sector is
\begin{equation}
\rho_A(q)=\frac{\Pi_{q}\, \rho_A\, \Pi_{q}}{p(q)}.
\end{equation}
The \textit{symmetry-resolved entanglement entropy} in this sector is then defined as
\begin{equation}\label{SREdef}
S_1(q)=-\mathrm{Tr}\!\left[ \rho_A(q) \ln \rho_A(q) \right].
\end{equation}
To compute the symmetry-resolved entropies, one must determine the entanglement spectrum of $\rho_A$ and resolve it into contributions from different charge sectors. This procedure is highly nontrivial, and in most cases poses significant difficulties for analytic calculations. A convenient tool to access these quantities is the \textit{charged moments} \cite{Goldstein:2017bua}
\begin{equation}\label{znalpha}
Z_n(\alpha) = \mathrm{Tr}\!\left[\rho_A^n e^{i \alpha Q_A}\right],
\end{equation}
where $n$ is the Rényi index and $\alpha$ plays the role of a fictitious chemical potential conjugate to the charge. Fourier transformation with respect to $\alpha$ implements the projector $\Pi_{q}$, allowing one to extract the symmetry resolved moments \cite{Goldstein:2017bua,Xavier:2018kqb}
\begin{equation}\label{znq}
\mathcal{Z}_n(q) = \int_{-\pi}^{\pi} \frac{d\alpha}{2\pi}\, e^{-i\alpha q} Z_n(\alpha).
\end{equation}
The symmetry-resolved Rényi entropies are then given by
\begin{equation}\label{snq}
S_n(q)=\frac{1}{1-n} \ln \!\left( \frac{\mathcal{Z}_n(q)}{\mathcal{Z}^n_1(q)} \right),
\end{equation}
and taking the limit $n \to 1$ yields the symmetry-resolved von Neumann entropy $S_1(q)$. In addition, one observes that the probability distribution $p(q)$ of the charge sectors can be obtained via the moments as
\begin{equation}
p(q)=\mathcal{Z}_1(q).
\end{equation}
Moreover, using Eqs. \eqref{rhoAdecom} and \eqref{SREdef} the total entanglement entropy associated to $\rho_A$ can be decomposed as
\begin{eqnarray}\label{EEdecompose}
S_{E}=\sum_q p(q)S_1(q)-\sum_q p(q)\log p(q)\equiv S_c+S_f.
\end{eqnarray}
The first term represents the \textit{configurational entropy}, i.e. the average entanglement over charge sectors, while the second quantifies charge fluctuations across the bipartition and is referred to as the \textit{fluctuation (number) entropy}. In Ref. \cite{Lukin:2019}, it was shown that in systems featuring both interactions and disorder, the dynamical behaviors of the configurational and fluctuation contributions to entanglement are markedly different. While the configurational entropy $S_c$ displays a slow logarithmic growth in time, the fluctuation entropy $S_f$ rapidly saturates to an asymptotic value. Furthermore, in the many-body localized phase, the fluctuation entropy obeys an area-law scaling. This behavior can be traced back to the fact that $S_f$ is governed by particle-number fluctuations across the bipartition, which are strongly suppressed due to localization and arise solely from tunneling processes across the boundary. 
Moreover, the configurational entropy can be interpreted as a measure of the operationally accessible entanglement entropy \cite{Wiseman:2003qje,Barghathi:2018idr}.

Previous studies of symmetry-resolved entanglement have shown that, in conformal field theories with a global $U(1)$ symmetry, conformal invariance enforces an equal distribution of entanglement among all charge sectors \cite{Xavier:2018kqb}. Understanding how this equipartition of entanglement extends to other models beyond conformal field theories has therefore attracted significant attention \cite{Bonsignori:2019naz,Murciano:2019wdl,Capizzi2020,Murciano:2020vgh,Capizzi2022}. A central objective of the present work is to investigate to what extent this equipartition property persists in non-relativistic systems, where Lorentz and conformal symmetries are absent. Let us also note that symmetry-resolved entanglement has been extended to broad classes of quantum field theories beyond relativistic conformal settings. In particular, \cite{Pirmoradian:2023uvt} investigated symmetry-resolved entanglement entropy in non-local quantum field theories, highlighting qualitative differences induced by non-locality. More recently, symmetry resolution in explicitly non-Lorentzian field theories was systematically studied in \cite{Banerjee:2024ldl}, where the role of anisotropic scaling and non-relativistic symmetries in shaping the structure of entanglement across charge sectors was elucidated. Symmetry-resolved entanglement has also been examined in Schrödinger fermionic field theories, which display distinctive entanglement features driven by the anisotropic scaling symmetry \cite{Gentile:2025nwv}.

The paper is organized as follows. In section \ref{sec:setup} we introduce the non-relativistic models studied in this work, including the complex Lifshitz harmonic chain and the Dirac–Lifshitz fermion theory. In section \ref{sec:Lifchain} we investigate symmetry-resolved entanglement in the complex Lifshitz harmonic chain, first reviewing the framework of charged moments and symmetry resolution and then presenting the results for symmetry-resolved R\'enyi and von Neumann entropies. Section \ref{sec:liffermion} is devoted to symmetry-resolved entanglement in Lifshitz fermionic theories, where qualitative differences with bosonic models are discussed. Finally, section \ref{sec:conclu} contains our conclusions and a discussion of the physical implications of our results, along with possible directions for future work.

\section{Set-up}\label{sec:setup}

Our aim in this paper is to explore the symmetry resolution of the entanglement in non-relativistic quantum field theories. Specifically, we will focus on the so-called charged moments and symmetry resolved entanglement entropy. This study is devoted to the vacuum state in two spacetime dimensions. In this section, before turning to the details of the symmetry resolution, we begin with a brief review of our nonrelativistic models. We also provide rigorous definitions of all relevant quantities and a concise description of the computation scheme.

\subsection{The complex Lifshitz harmonic chain}

The first model we consider is a free complex scalar field theory in $1+1$ dimensions whose action is given by
\begin{eqnarray}\label{actionscalar}
\mathcal{S}_b=\int dt\;dx\left(\partial_t\phi^\dagger(x)\partial_t\phi(x)-\partial_x^z\phi^\dagger(x)\partial_x^z\phi(x)-m^{2z}\phi^\dagger(x)\phi(x)\right),
\end{eqnarray}
which is a generalization of relativistic complex Klein–Gordon theory and respects Lifshitz scaling symmetry
\begin{eqnarray}\label{Lifscaling}
t\rightarrow \lambda^z t,\;\;\;\;\;\;\;\;\;\;x\rightarrow \lambda x,
\end{eqnarray}
in the massless limit. Clearly, the above action exhibits a global $U(1)$ symmetry, \textit{i.e.}, a symmetry under phase transformations of the field given by $\phi(x)\rightarrow e^{i\theta} \phi(x)$. Moreover, the Hamiltonian of this theory is
\begin{eqnarray}\label{hamil}
H=\int dx\left(\pi^\dagger(x)\pi(x)+\partial_x^z\phi^\dagger(x)\partial_x^z\phi(x)+m^{2z}\phi^\dagger(x)\phi(x)\right),
\end{eqnarray}
which is still second order in time derivative and thus the corresponding conjugate momentum is $\pi(x)=\partial_t\phi^\dagger(x)$ as expected. The field operator can be written in terms of the particles and antiparticles modes as follows
\begin{eqnarray}\label{phi}
\phi(x)=\int\frac{dk}{2\pi}\frac{1}{\sqrt{2\omega_k}}\left(a_ke^{ikx}+b_k^\dagger e^{-ikx}\right),
\end{eqnarray}
where $\omega_k^2=k^{2z}+m^{2z}$ is the dispersion relation in the continuum limit. Moreover, the Hamiltonian becomes 
\begin{eqnarray}\label{hami2}
H=\int \frac{dk}{2\pi} \omega_k\left(a^\dagger_k a_k+b^\dagger_k b_k\right).
\end{eqnarray}
Further, by Noether's theorem, the $U(1)$ symmetry leads to a conserved quantity as follows
\begin{eqnarray}\label{Q}
Q=\int \frac{dk}{2\pi} \left(a^\dagger_k a_k-b^\dagger_k b_k\right).
\end{eqnarray}
Interestingly, the theory introduced in Eq. \eqref{hamil} can be rewritten in terms of two real fields and their conjugate momenta. To do so we define
\begin{eqnarray}\label{hamil}
\pi(x)=\frac{1}{\sqrt{2}}\left(\pi_1(x)+i\pi_2(x)\right),\hspace*{2cm}\phi(x)=\frac{1}{\sqrt{2}}\left(\phi_1(x)+i\phi_2(x)\right)
\end{eqnarray}
which yields
\begin{eqnarray}\label{hamil1}
H=\sum_{j=1, 2}\int dx\left(\pi_j^2(x)+\partial_x^z\phi_j(x)\partial_x^z\phi_j(x)+m^{2z}\phi_j^2(x)\right)\equiv \sum_{j=1, 2} H_{\mathbb{R}}(\phi_j),
\end{eqnarray}
where each $H_{\mathbb{R}}$ denotes the Hamiltonian for a real scalar field. This result allows us to adopt the established framework for computing entanglement measures in a real harmonic chain. In the following sections, we evaluate these entanglement measures using the correlator method, an efficient algorithm that computes the eigenvalues of the reduced density matrix for Gaussian states (e.g., see \cite{Peschel:2002yqj,Peschel:2004qbn,Eisler:2009vye}). To implement this, we first discretize the model on a one-dimensional lattice with periodic boundary conditions and a finite lattice spacing $\epsilon$. In this discretized setup, the corresponding Hamiltonian can be written as \cite{MohammadiMozaffar:2017nri}
\begin{eqnarray}\label{hamil1dis}
H=\frac{1}{2}\sum_{n=1}^N\left(p_n^2+(\sum_{j}{_z}c_j\;q_{n+j-1})^2+m^{2z}q_n^2\right),
\end{eqnarray}

where ${_z}c_j=(-1)^{z+j}\frac{z!}{j!(z-j)!}$, and, without loss of generality, we set the lattice spacing to unity. Moreover, upon transforming to the momentum basis, the above Hamiltonian can be rewritten in the following form
\begin{eqnarray}
H=\int dk\left[a_k^\dagger a_k+a_{-k}^\dagger a_{-k}\right]\omega_k,
\end{eqnarray}
where $\omega_k=\sqrt{m^{2z}+\left(2\sin\frac{k}{2}\right)^{2z}}$. Now, we construct the following two-point correlation functions
\begin{eqnarray}
X_{mn}=\frac{1}{2\pi}\int dk\;\omega_k^{-1}\;e^{i(m-n)},\hspace{1cm}P_{mn}=\frac{1}{2\pi}\int dk\;\omega_k^\;\;e^{i(m-n)},
\end{eqnarray}
where $m, n\in [1, \ell]$ and $\ell$ denotes the number of lattice sites contained within $A$. Next, having the eigenvalues $\nu_k$ of $\sqrt{X.P}$, it is then possible to determine the behavior of
the several entanglement measures including R\'enyi and entanglement entropies. For example, in the case of the entanglement entropy, the corresponding expression takes the following form
\begin{eqnarray}\label{SEE}
S_E=\sum_{k=1}^{\ell}\left(\left(\nu_k+\frac{1}{2}\right)\log\left(\nu_k+\frac{1}{2}\right)-\left(\nu_k-\frac{1}{2}\right)\log\left(\nu_k-\frac{1}{2}\right)\right).
\end{eqnarray}
Further, as shown in \cite{Murciano:2019wdl}, an exact correlation matrix approach to symmetry-resolved entanglement was explicitly developed for the complex harmonic chain. In particular, by constructing the corresponding reduced density matrix of a subsystem of length $\ell$ and employing the definition in Eq.~\eqref{znalpha}, the charged moments can be expressed as
\begin{eqnarray}\label{Zscalar}
Z_n(\alpha)=e^{F_n(\alpha)+F_n(-\alpha)}, \hspace{1cm}\text{with}\hspace{1cm}F_n(\alpha)=\log\prod_{k=1}^{\ell}\frac{1}{(\nu_k+\frac{1}{2})^n-e^{i\alpha}(\nu_k-\frac{1}{2})^n}.
\end{eqnarray}
Inserting the above expression into Eq.~\eqref{znq} and using Eq.~\eqref{snq}, one can find the symmetry resolved entropies. By employing this approach, various features of symmetry-resolved entanglement measures in scalar quantum field theories have been extensively explored, unveiling a rich and diverse structure of results and physical interpretations, e.g., \cite{Murciano:2019wdl,Murciano:2020vgh}. In the following section, we generalize these investigations to a Lifshitz scalar field theory, aiming to explore the impact of a nontrivial dynamical exponent $z$ on the scaling properties of the symmetry-resolved entanglement measures.

\subsection{The Dirac-Lifshitz fermion theory}
As the second model, we consider a free fermion theory in two dimensions, characterized by the following action \cite{Vasli:2024mrf}
\begin{eqnarray}\label{actionfermion}
\mathcal{S}_f=\int dt\;dx\;\bar{\Psi}\left(i\gamma^t\partial_t+i\gamma^x\mathcal{D}^{z-1}\partial_x-m^{z}\right)\Psi,
\end{eqnarray}
where $\mathcal{D}=\sqrt{-\partial^x\partial_x}$. Again, the Lifshitz scaling symmetry emerges in the massless limit. To proceed, we follow the same steps as in the scalar case and place the theory on a one-dimensional lattice. The corresponding discretized Hamiltonian then takes the form 
\begin{eqnarray}\label{Hdirac}
H_f=\sum_{n=1}^{N}\bar{\Psi}_n\left(-\gamma^x f(k)+m^z\right)\Psi_n,
\end{eqnarray}
where for odd and even values of the dynamical exponent $z$ we have
\begin{eqnarray}\label{f}
f_{\text{odd}}(k)=-\left(\sin k\right)^z,\hspace{2cm}f_{\text{even}}(k)=|\sin k|\,\left(\sin k\right)^{z-1}.
\end{eqnarray}
Furthermore, upon transforming to the momentum basis, the Hamiltonian takes the form
\begin{eqnarray}\label{Hamildissdirac1}
H_f=\int_{-\pi}^{\pi}dk\,\left(a^\dagger_{-k}a_{-k}+a^\dagger_{k}a_{k}\right)\omega_{k},
\end{eqnarray}
where $\omega_{k}=\sqrt{f^2(k)+m^{2z}}$. We again utilize the correlator approach to compute the reduced density matrix spectrum and the associated entanglement measures. In particular, by defining the corresponding two-point correlation function as $C_{rs}\equiv\langle \Psi_r^\dagger \Psi_s \rangle$ we obtain
\begin{eqnarray}\label{C}
C_{rs}=\frac{\delta_{rs}}{2}\textbf{1}_{2\times 2}-\frac{1}{4\pi}\int_{-\pi}^{\pi}dk
\left(\begin{matrix}
f(k) & m^z\\
m^z & -f(k)
\end{matrix}\right) \frac{e^{i(r-s)k}}{\omega_k}.
\end{eqnarray}
Subsequently, various entanglement measures can be expressed in terms of the spectrum of the above correlation matrix. In particular, the entanglement entropy is given by 
\begin{eqnarray}
S_E=-\sum_{k=1}^{\ell}\left(\nu_k\log\nu_k+\left(1-\nu_k\right)\log\left(1-\nu_k\right)\right).
\end{eqnarray}
Furthermore, one can also determine the symmetry-resolved entanglement measures for fermionic lattice chains from the eigenvalues of the correlation matrix. In particular, the charged moments take the form \cite{Goldstein:2017bua,Murciano:2020vgh}\footnote{At half-filling, particle-hole symmetry pairs the eigenvalues $\nu_k$ such that Eq. \eqref{Znfermion1} can be rewritten as 
\begin{eqnarray}\label{Znfermion}
Z_n(\alpha)=\prod_{k=1}^{\ell}\left(\nu_k^n e^{i\frac{\alpha}{2}}+\left(1-\nu_k\right)^n e^{-i\frac{\alpha}{2}}\right)=e^{i \alpha \ell / 2}g_n(\alpha).
\end{eqnarray}
The overall phase $e^{i \alpha \ell / 2}$ accounts for the mean charge in the subsystem, while the function $g_n(\alpha)$ is real and even in $\alpha$ \cite{Bonsignori:2019naz}.
}
\begin{eqnarray}\label{Znfermion1}
Z_n(\alpha)=\prod_{k=1}^{\ell}\left(\nu_k^n e^{i\alpha}+\left(1-\nu_k\right)^n\right),
\end{eqnarray}
By substituting the result into Eq.~\eqref{znq} and using Eq.~\eqref{snq}, we obtain the symmetry resolved entropies. Recent progress has extended the study of symmetry-resolved entanglement to fermionic systems, revealing detailed information about charge fluctuations and sector-dependent correlations. The first comprehensive analysis appeared in \cite{Bonsignori:2019naz}, where exact methods were developed to compute charge-resolved Rényi entropies in free fermionic chains. Subsequent investigations explored long-range models \cite{Ares:2022hdh}, and critical scaling behavior in conformal systems \cite{Jones:2022tgp}. Finite-size and thermal effects in continuum Dirac fermion theories were also addressed in \cite{Foligno:2022ltu}, further enriching our understanding of symmetry resolution in fermionic quantum field theories. In what follows, we generalize the framework to a Lifshitz Dirac field theory and analyze how a nontrivial dynamical exponent $z$ modifies the scaling structure of the symmetry-resolved entanglement.

\section{Symmetry resolved entanglement in complex Lifshitz harmonic chain}\label{sec:Lifchain}

To begin exploring symmetry resolution in Lifshitz field theories, we consider the vacuum state of a scalar model. Our analysis starts with the charged moments, which provide a convenient tool to probe the distribution of entanglement across different charge sectors. Building on the insights gained from these quantities, we then generalize the computation to obtain the symmetry-resolved Rényi and von Neumann entropies.

\subsection{Charged moments and symmetry resolution}

\begin{figure}[h!]
	\begin{center}
\includegraphics[scale=0.78]{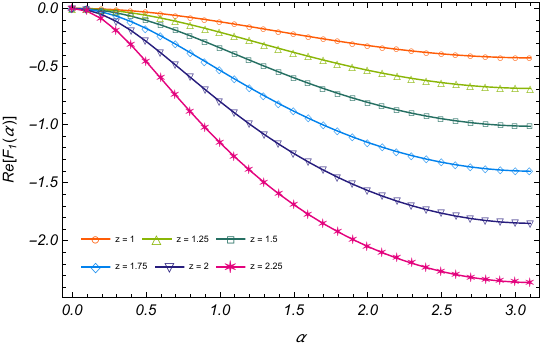}
  \hspace*{0.4cm}
\includegraphics[scale=0.78]{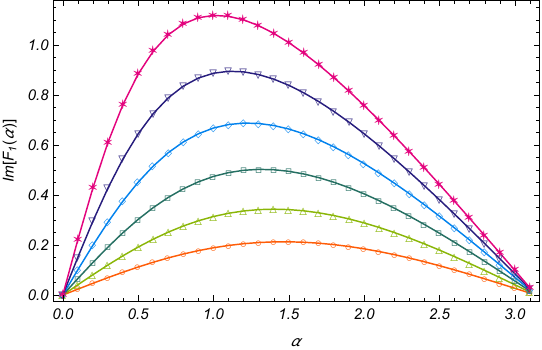}
\includegraphics[scale=0.78]{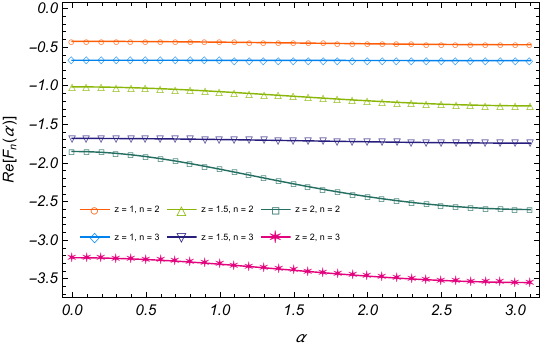}
  \hspace*{0.4cm}
\includegraphics[scale=0.78]{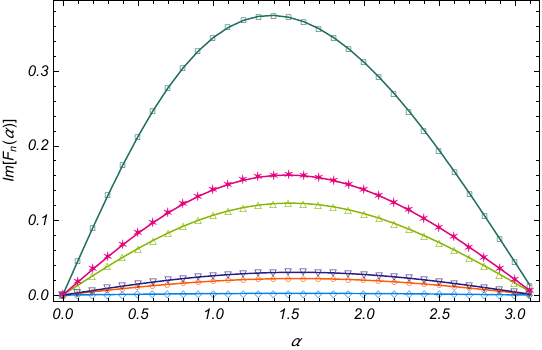}
  	\end{center}
	\caption{ The real (left) and the imaginary (right) parts of the charged moments as functions of $\alpha$ for several values of $z$ and $n$ with $m=0.1$ and $\ell=50$. }
	\label{fig:ReF1alphaz}
\end{figure}

\begin{figure}[h!]
	\begin{center}
\includegraphics[scale=0.78]{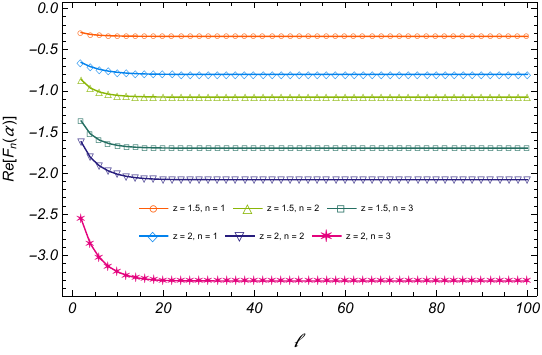}
  \hspace*{0.4cm}
\includegraphics[scale=0.78]{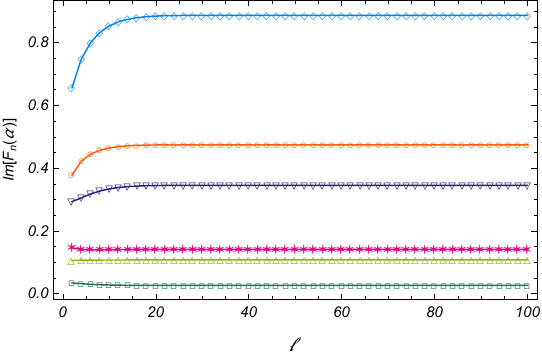}
  	\end{center}
	\caption{ The real (left) and the imaginary (right) parts of the charged moments as functions of $\ell$ for several values of $z$ and $n$ with $m=0.1$ and $\alpha=1$. }
	\label{fig:ReFnalphaellz}
\end{figure}

\begin{figure}[h!]
	\begin{center}
\includegraphics[scale=0.58]{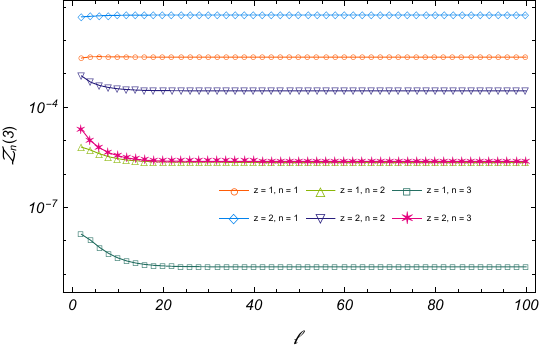}
  \includegraphics[scale=0.58]{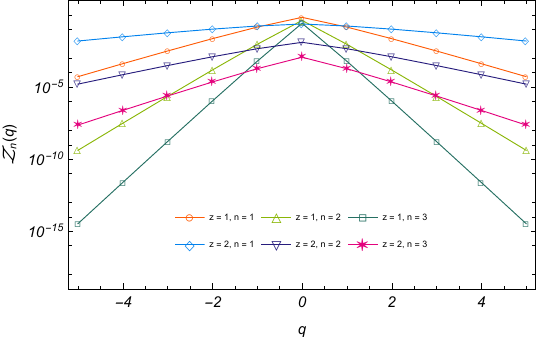}
  \includegraphics[scale=0.58]{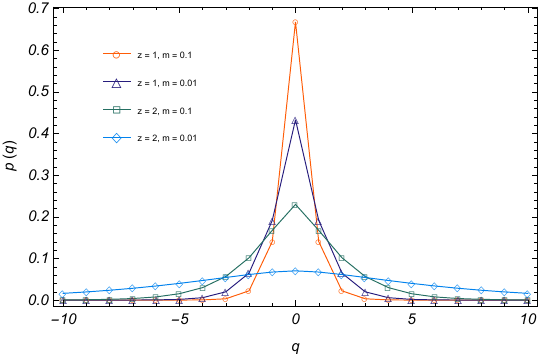}
  	\end{center}
	\caption{ \textit{Left}: Symmetry resolved partition sums as function of the width of the entangling region for several values of the dynamical exponent and R\'enyi index with $q=3$. \textit{Middle}: The same quantity as function of charge for the same values of the parameters with $\ell=50$. In both panels we set $m=0.1$. \textit{Right}: The probability of measuring charge $q$ within the subsystem for different $z$ and $m$. }
	\label{fig:Zmathcal}
\end{figure}

We begin by examining the charged moments for an interval of length $\ell$ in an infinite harmonic chain, as defined in Eq.~\eqref{Zscalar}. The numerical results are shown in figures \ref{fig:ReF1alphaz}, \ref{fig:ReFnalphaellz} and \ref{fig:Zmathcal} for various parameter values. In the following, we focus primarily on $N=2000$, since the qualitative features of the symmetry-resolved measures are largely independent of the total system size. Figure~\ref{fig:ReF1alphaz} shows the real and imaginary parts of $F_n(\alpha)$ as functions of $\alpha$ for various values of the dynamical exponent, with fixed parameters $m = 0.1$ and $\ell = 50$. The first row corresponds to the case $n = 1$. In particular, the left panel displays the real part of $F_1(\alpha)$, as defined in Eq.~\eqref{Zscalar}, which exhibits a monotonically decreasing behavior with increasing $\alpha$ and $z$. Considering now the charged moments of the complex chain, defined by $\log Z_n(\alpha) = 2\,\mathrm{Re}\,F_n(\alpha)$, we observe a similar scaling behavior. On the other hand, the imaginary part, shown in the right panel, exhibits a distinctly different behavior. Here, although $\mathrm{Im}\,F_n(\alpha)$ exhibits a non-monotonic dependence on $\alpha$, it increases monotonically with the dynamical exponent. These behaviors are qualitatively similar to the scaling of the $F_n(\alpha)$ with mass in the relativistic case with $z=1$. Indeed, from the Lifshitz dispersion relation, we observe that the spectrum of the reduced density matrix depends on $m^{z}$. Consequently, for larger values of the dynamical exponent, the effective mass of the theory decreases, and the previously observed scaling behavior of the charged moments in the relativistic setup is recovered.\footnote{As mentioned earlier, we set the lattice spacing to unity. Therefore, to maintain a consistent truncation from lattice results to the continuum, we focus on masses $m < 1$. In our numerics, we primarily consider $m = 0.1$, since the qualitative features of the scaling and behavior of symmetry-resolved quantities are independent of the mass parameter in this regime.} The plots in the second row correspond to $n > 1$, where similar qualitative features are observed.

In order to gain further insights into certain properties of $F_n(\alpha)$ we summarize the full $\ell$-dependence of this quantity in figure \ref{fig:ReFnalphaellz} for $\alpha=1$. The numerical results show that the charged moments first decrease very sharply with $\ell$ and then suddenly saturates to a constant value. Further, the saturation value for the real part is a monotonically decreasing function of $z$. It is worth to mention that considering other values of $\alpha$ the qualitative features of these results do not change.

In figure \ref{fig:Zmathcal}, we present the numerically computed symmetry-resolved moments as functions of the subsystem size and charge. In the left panel, $\mathcal{Z}_n(q)$ is plotted as a function of $\ell$ for several values of the dynamical exponent and R\'enyi index, with the charge fixed at $q = 3$. We see that the moments decrease with the subsystem size and then saturates to a constant value. Moreover, the saturation value is a monotonically increasing function of $z$ due to the enhancement of the correlation between the lattice sites. The middle panel of figure \ref{fig:Zmathcal} illustrates the symmetry-resolved moments as a function of $q$ for $\ell=50$. Our numerical results indicate that the variance of $\mathcal{Z}_n(q)$ increases with the dynamical exponent for any fixed $n$. Hence, we expect that increasing $z$ enhances the contribution of sectors with larger charges to the symmetry-resolved measures. In other words, as the dynamical exponent grows, the charge distribution broadens, leading to a more significant weight of highly charged sectors in the total entanglement. This behavior suggests that the interplay between the dynamical scaling and charge fluctuations becomes increasingly relevant in the nonrelativistic regime. The results displayed in the right panel further support this conclusion. In this plot, we show $p(q)$, the probability of measuring charge $q$ within the subsystem, for various values of $z$ and $m$. Once again, we observe that increasing the dynamical exponent produces the same effect as reducing the mass. We will examine these features in more detail in the following subsection. 

\subsection{Symmetry resolved R\'enyi and entanglement entropies}
In this subsection we explicitly consider entanglement and R\'enyi entropies and work out their symmetry resolution. In the following, we focus on the symmetry-resolved measures for $z > 1$ and compare them with the relativistic case to highlight the distinctive physical signatures of the nonrelativistic regime.

\begin{figure}[h!]
	\begin{center}
\includegraphics[scale=0.78]{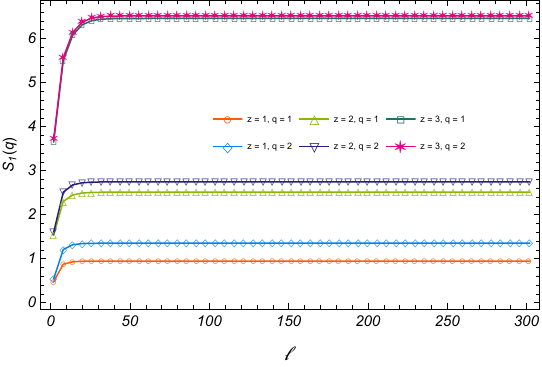}
  \hspace*{0.4cm}
\includegraphics[scale=0.78]{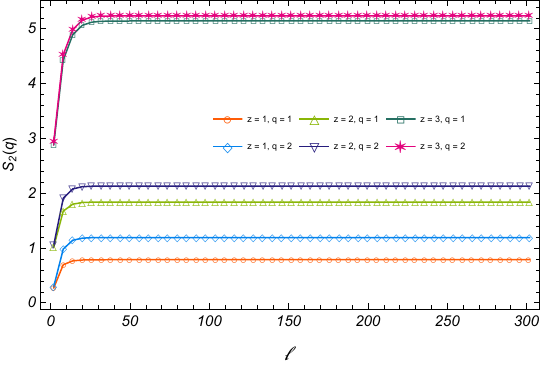}
  	\end{center}
	\caption{Symmetry resolved R\'enyi entropies as the function of the entangling width for several values of the dynamical exponent and charge with $n=1$ (left) and $n=2$ (right). In both panels we we set $m=0.1$. For larger values of the dynamical exponent we have an approximate equipartition.}
	\label{fig:Snellq}
\end{figure}

Figure \ref{fig:Snellq} displays the symmetry-resolved entanglement and Rényi entropies as functions of the subsystem size $\ell$ for several values of the dynamical exponent $z$ and charge $q$, with the mass fixed at $m=0.1$. The left and right panels correspond to the cases $n=1$ and $n=2$, respectively.
A few salient features can be observed. First, both entanglement measures increase with the dynamical exponent. Second, while for small $z$ the symmetry-resolved entropies exhibit a pronounced dependence on $q$, this dependence becomes much weaker for larger $z$, approaching an approximate equipartition among charge sectors. Moreover, increasing the Rényi index~ $n$ tends to suppress this equipartition, and achieving a clearer equal distribution requires considering larger values of the dynamical exponent.

\begin{figure}[h!]
	\begin{center}
\includegraphics[scale=0.78]{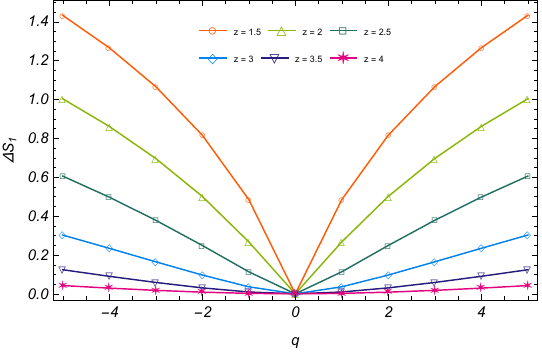}
  \hspace*{0.4cm}
\includegraphics[scale=0.78]{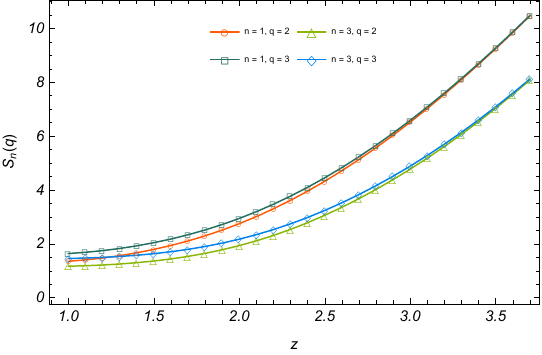}
  	\end{center}
	\caption{\textit{Left}: Subtracted symmetry resolved entanglement entropy as function of charge for several values of the dynamical exponent. \textit{Right}: Symmetry resolved R\'enyi entropies as function of the dynamical exponent for several values of charge and R\'enyi index. In both panels we set $m=0.1$ and $\ell=50$. For larger values of the dynamical exponent we have an approximate equipartition.}
	\label{fig:deltaSnellq}
\end{figure}

To further elucidate the dependence of the symmetry-resolved entanglement measures on the charge and the dynamical exponent, we summarize our numerical results in figure \ref{fig:deltaSnellq} for a fixed subregion length. The left panel displays the subtracted symmetry-resolved entanglement entropy, defined as $\Delta S_1(q)\equiv S_1(q)-S_1(0)$ plotted as a function of the charge $q$ for several values of the dynamical exponent $z$. As is evident from the figure, $\Delta S_1(q)$ is an increasing function of the charge. Moreover, for small values of $z$, this variation with $q$ is pronounced, indicating a strong charge dependence of the entanglement contribution. However, as the dynamical exponent increases, $\Delta S_1(q)$ gradually approaches zero, suggesting that the entanglement becomes uniformly distributed among the charge sectors. This behavior signals the emergence of an \textit{effective equipartition} of entanglement in the large-$z$ regime. 

\begin{figure}[h!]
	\begin{center}
\includegraphics[scale=1]{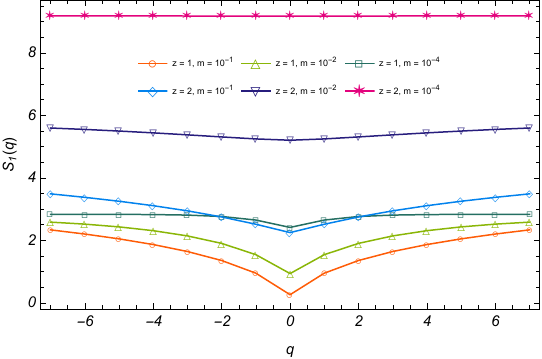}
  	\end{center}
	\caption{Symmetry resolved entanglement entropy as function of charge for several values of mass and dynamical exponent with $\ell=50$. For larger (smaller) values of the dynamical exponent (mass) the equipartition recovered.}
	\label{fig:S1qz}
\end{figure}

Furthermore, the right panel of Fig. \ref{fig:deltaSnellq} illustrates the behavior of the symmetry-resolved R\'enyi entropies as functions of the dynamical exponent $z$ for several fixed values of the charge $q$ and R\'enyi index $n$. Consistent with the trends observed for the entanglement entropy, we find that at small values of $z$, the symmetry-resolved R\'enyi entropies exhibit a noticeable dependence on the charge sector, reflecting a nonuniform distribution of entanglement among the different $U(1)$ sectors. However, as $z$ increases, this dependence gradually weakens, and the different charge sectors tend to contribute equally to the total R\'enyi entropy. This behavior clearly indicates the emergence of an approximate equipartition of entanglement in the large-$z$ limit, in full agreement with the results discussed previously for the charged moments and the symmetry-resolved entanglement entropy. 
Interestingly, this behavior becomes more pronounced as the mass parameter decreases. An explicit illustration of this behavior is presented in Fig.~\ref{fig:S1qz}, where one can clearly observe that approaching the Lifshitz fixed point enhances the degree of equipartition among the charge sectors. In particular, as the dynamical exponent increases, the symmetry-resolved entanglement entropy tends toward an almost uniform distribution, signaling a stronger manifestation of equipartition in the critical regime. It is worth noting that for the relativistic critical scalar ($z=1$, $m=0$), equipartition holds to all orders in the UV cutoff as a consequence of representation theory \cite{DiGiulio:2022jjd,Northe:2023khz}. To examine whether a similar property extends to our non-relativistic setup, we have tested the UV cutoff dependence by varying $\ell$ with fixed $m\ell$ and $\ell/N$. Within numerical precision, the equipartition behavior remains unchanged, indicating that it is not an artifact of the UV cutoff.

\begin{figure}[h!]
	\begin{center}
\includegraphics[scale=0.8]{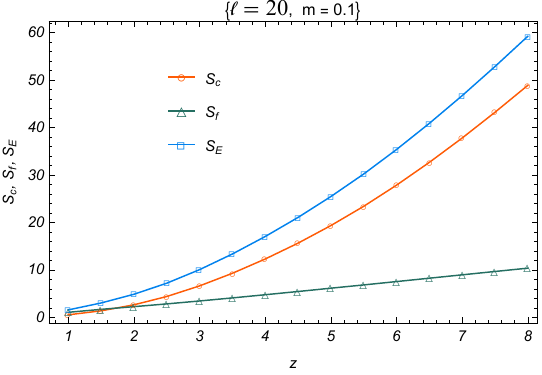}
\hspace*{1cm}
\includegraphics[scale=0.8]{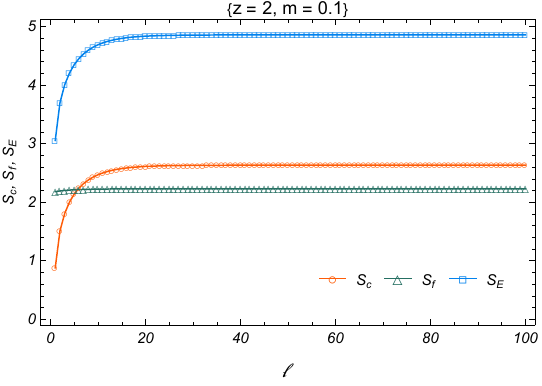}
  	\end{center}
	\caption{Configurational, fluctuation and total entanglement entropies as functions of the dynamical exponent (left) and the length of the entangling region (right).}
	\label{fig:SfScStotz}
\end{figure}

To conclude this section, we comment on the behavior of the configurational and fluctuation entropies in the nonrelativistic regime. Figure \ref{fig:SfScStotz} shows the dependence of these quantities on the dynamical exponent and the length of the entangling region for various parameter choices. For completeness, we also include the total entanglement entropy, consistent with Eq.~\eqref{EEdecompose}. A key observation is that, for small values of $z$, the fluctuation entropy $S_f$ exceeds the configurational entropy $S_c$, whereas for larger $z$, this relation is reversed, yielding $S_c > S_f$. Both $S_c$ and $S_f$ increase monotonically with the dynamical exponent; however, only $S_c$ continues to grow without bound, while $S_f$ exhibiting an approximately linear scaling at large $z$. Consequently, in this limit, $S_f \ll S_c$, indicating that the contribution from charge fluctuations within the subsystem to the total entanglement entropy becomes negligible as the Lifshitz exponent increases. Interestingly, as discussed in \cite{Wiseman:2003qje,Barghathi:2018idr}, the configurational entropy corresponds to the operationally accessible part of the entanglement entropy. Therefore, we expect $S_c$ to become increasingly significant in nonrelativistic settings — a point that deserves further investigation. In addition, the right panel demonstrates that in the limit of small subsystem length, charge fluctuations dominate the total entanglement entropy, whereas the configurational entropy, which corresponds to the operationally accessible portion, is significantly reduced.

\section{Symmetry resolved entanglement in Lifshitz fermion theories}\label{sec:liffermion}

In this section we advance our analysis by explicitly computing the symmetry-resolved entropies for the Lifshitz fermion theory defined in Eq. \eqref{actionfermion}. As shown in \cite{Vasli:2024mrf}, the entanglement entropy in the vacuum of a Lifshitz fermion theory depends sensitively on the value of the mass parameter. Motivated by this observation, our investigation proceeds along two complementary directions. First, we perform a detailed numerical study of the symmetry-resolved entanglement measures at finite mass. By scanning a range of parameter values, we characterize their behavior across different regimes and identify the associated scaling features. This analysis allows us to directly probe how non-relativistic Lifshitz scaling influences the structure of charge-resolved entanglement, and in particular how departures from relativistic scaling symmetry manifest within each charge sector. Second, we turn to the small-mass regime, where we supplement the numerical findings with analytic expressions obtained through controlled expansions near zero mass. These analytic results clarify the universal properties of the symmetry-resolved quantities in the massless limit and illuminate their asymptotic behavior. In addition, they provide a useful benchmark for interpreting and validating the numerical data. Taken together, these two approaches yield a coherent and detailed understanding of symmetry-resolved entanglement in Lifshitz fermion systems, highlighting the interplay between non-relativistic scaling, mass deformations, and the structure of charge sectors.

\subsection{From the charged moments to symmetry resolution}

\begin{figure}[h!]
	\begin{center}
     \includegraphics[scale=0.8]{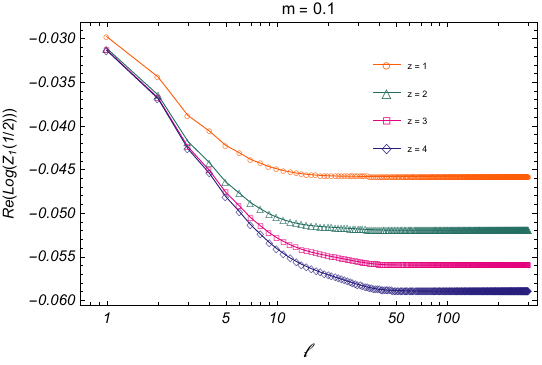}
     \hspace*{1cm}
\includegraphics[scale=0.8]{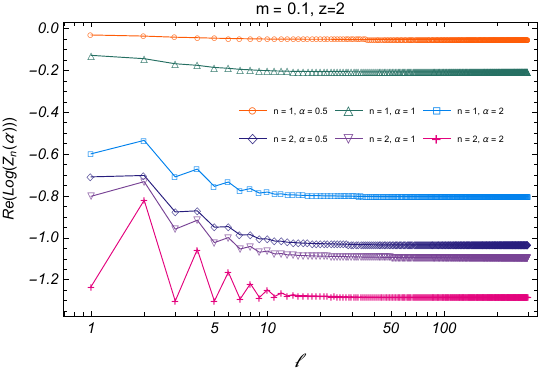}
  	\end{center}
	\caption{Real part of the logarithm of the charged moments as a function of $\ell$ for several values of the parameters in fermionic theory.}
	\label{fig:Znzellfermion}
\end{figure}

\begin{figure}[h!]
	\begin{center}
\includegraphics[scale=0.8]{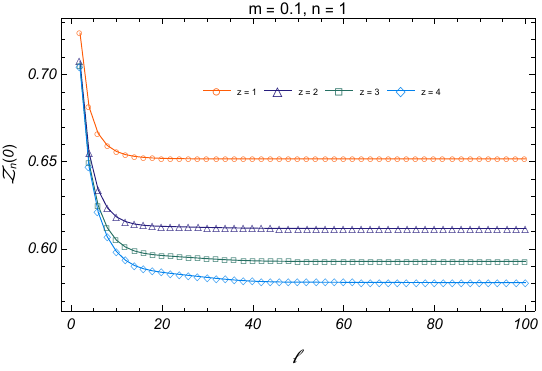}
\hspace*{1cm}
\includegraphics[scale=0.8]{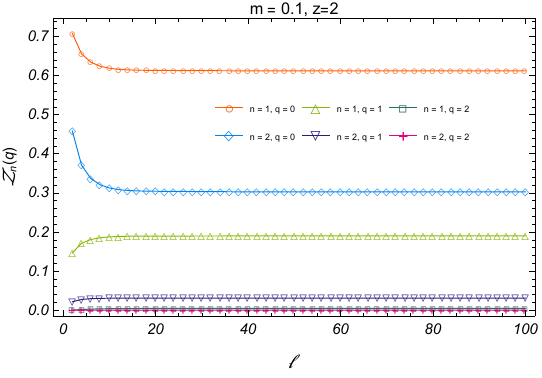}
  	\end{center}
	\caption{ \textit{Left}: Symmetry resolved partition sums as function of the width of the entangling region for several values of the dynamical exponent in fermion theory. \textit{Right}: The same quantity for different values of the R\'enyi index and charge with fixed dynamical exponent.}
	\label{fig:Zmathcalfermionell}
\end{figure}

\begin{figure}[h!]
	\begin{center}
\includegraphics[scale=0.8]{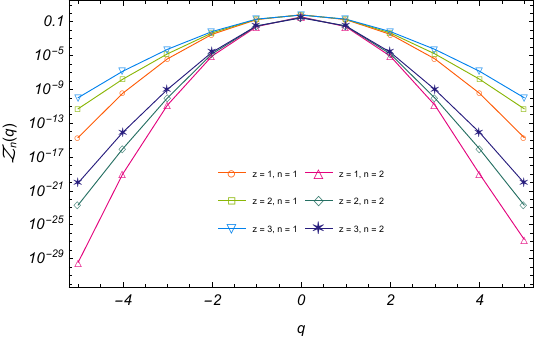}
\hspace*{1cm}
\includegraphics[scale=0.8]{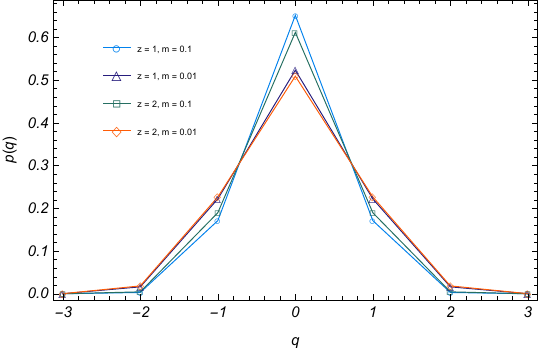}
  	\end{center}
	\caption{ \textit{Left}: Symmetry resolved partition sums as function of charge with $\ell=50$ and $m=0.1$. \textit{Right}: The probability of measuring charge $q$ within the subsystem for different values of $z$ and $m$.}
	\label{fig:Zmathcalfermion}
\end{figure}

We begin by analyzing the charged moments and the symmetry-resolved partition sums for a range of values of the mass parameter $m$, flux $\alpha$, and the R\'enyi index $n$. Our starting point is the set of expressions for these quantities provided in Eqs.~\eqref{znq} and \eqref{snq}, which allow us to numerically evaluate the symmetry-resolved measures across different regions of parameter space. This enables us to track how each symmetry sector responds to changes in mass and to investigate how non-relativistic Lifshitz scaling influences the detailed structure of symmetry resolution. The resulting numerical data are presented in figures~\ref{fig:Znzellfermion}, \ref{fig:Zmathcalfermionell} and \ref{fig:Zmathcalfermion}. These figures display the behavior of the various symmetry-resolved quantities across different parameter regimes.

In the left panel of figure \ref{fig:Znzellfermion} we present the numerical results for the charged moments in the presence of a flux insertion $\alpha$, evaluated for several values of the dynamical exponent $z$ and mass $m$ at fixed R\'enyi index $n=1$. The data show a clear trend: as the dynamical exponent increases, the real part of the charged moments systematically decreases. Moreover, in the large-subregion limit, the charged moments saturate to a constant value whose magnitude increases monotonically with the mass, indicating that massive excitations enhance the asymptotic contribution to the charged moments. The right panel of figure \ref{fig:Znzellfermion} displays the charged moments as functions of the subsystem size $\ell$ for a set of representative values of $(n,\alpha)$ with the dynamical exponent fixed to $z=2$. Increasing either the R\'enyi index or the flux parameter leads to a noticeable suppression in the magnitude of the charged moments across the entire range of $\ell$. A closer examination of the data reveals the presence of two distinct scaling regimes. In the small-$\ell$ regime, the charged moments exhibit pronounced oscillatory behavior superimposed on the leading trend, signaling the importance of subleading corrections that become particularly significant at small entangling lengths. As the subsystem size increases, these oscillations gradually diminish in amplitude and eventually become negligible. In the large-$\ell$ regime, the charged moments approach a smooth, nearly constant limiting behavior in which the oscillatory structure is strongly suppressed. Furthermore, we observe that the amplitude of the oscillations in the small-$\ell$ regime increases with both the R\'enyi index $n$ and the flux $\alpha$. This enhanced oscillatory response highlights the heightened sensitivity of the symmetry-resolved quantities to these parameters at short distances, where the interplay between flux, mass, and Lifshitz scaling is most pronounced.

Figure \ref{fig:Zmathcalfermionell} illustrates the behavior of $\mathcal{Z}_n(q)$ as a function of $\ell$ for several parameter choices. The left panel shows that the symmetry-resolved partition sums in the zero-charge limit are decreasing functions of the subsystem length, tending towards a constant value in the large-$\ell$ limit. In contrast to the bosonic case, this saturation value decreases monotonically with increasing $z$. The right panel presents the symmetry-resolved partition sums for several values of the R\'enyi index $n$ and the charge $q$. The numerical data reveal a clear qualitative distinction between the neutral and charged sectors. For the neutral sector ($q=0$), the symmetry-resolved partition sum begins at a non-zero value in the small-length limit and then decreases monotonically as $\ell$ increases, eventually saturating in the large-$\ell$ limit. This behavior reflects the progressive redistribution of weight among different charge sectors as the size of the entangling region grows. In contrast, for non-zero values of the charge, the corresponding quantities $\mathcal{Z}_n(q)$ increase monotonically with $\ell$. Moreover, as expected, they decrease with increasing charge.

Figure \ref{fig:Zmathcalfermion} displays the behavior of the symmetry-resolved partition sums as a function of charge for several parameter choices. In the left panel, illustrating the moments at a fixed subsystem length $\ell=50$, the numerical results indicate that the variance of $\mathcal{Z}_n(q)$ increases with the dynamical exponent for any fixed R\'enyi index. Consequently, similar to the scalar case, an increase in $z$ enhances the contribution of sectors with larger charges to the symmetry-resolved entropies. Finally, the right panel presents the probability distribution $p(q)$ of measuring a given charge within the subsystem for different values of $z$ and $m$. Here, we observe that increasing the dynamical exponent has the same qualitative effect as decreasing the mass.

\begin{figure}[h!]
	\begin{center}
\includegraphics[scale=0.78]{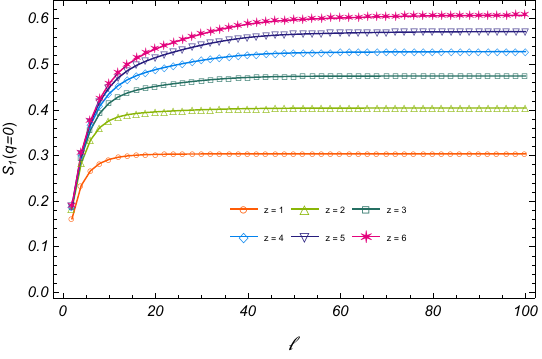}
  \hspace*{0.4cm}
\includegraphics[scale=0.78]{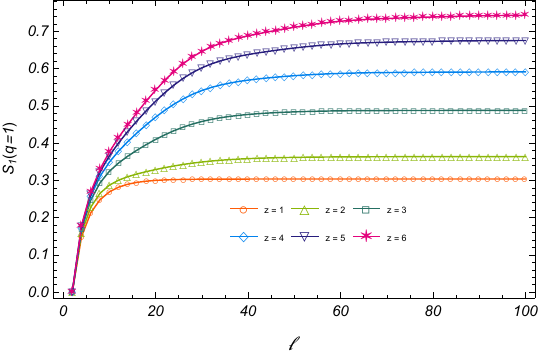}
  	\end{center}
	\caption{Symmetry resolved entanglement entropy as a function of the entangling length for several values of the dynamical exponent and charge in fermion theory with $m=0.1$.}
	\label{fig:Sqelln1fermion}
\end{figure}

\begin{figure}[h!]
	\begin{center}
\includegraphics[scale=0.8]{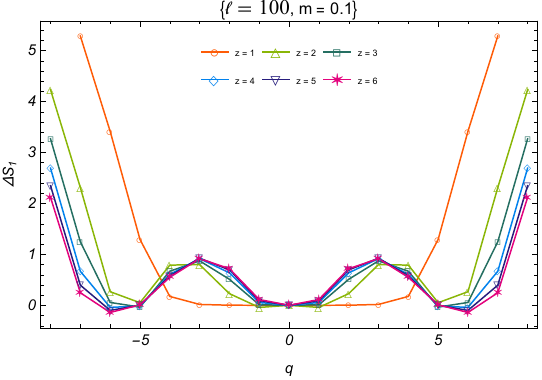}
  	\end{center}
	\caption{Symmetry resolved entanglement entropy as a function of charge for several values of the dynamical exponent.}
	\label{fig:S1qzfermion}
\end{figure}

Figure \ref{fig:Sqelln1fermion} presents the symmetry-resolved entanglement entropy $S_q(\ell)$ as a function of subsystem length for several values of the dynamical exponent, with the mass fixed at $m=0.1$. The left and right panels correspond to the cases of zero charge ($q=0$) and unit charge ($q=1$), respectively. We observe that $S_q(\ell)$ generally increases with the dynamical exponent. Furthermore, for smaller values of the charge and the dynamical exponent, the symmetry-resolved entropy exhibits a sharp saturation to a constant value. However, this transition becomes significantly smoother as $q$ and $z$ are increased.
To further clarify the dependence of the symmetry-resolved entanglement entropy on charge, we present $\Delta S_1(q)$ as a function of $q$ for several values of the dynamical exponent in Fig.~\ref{fig:S1qzfermion}. As the figure illustrates, genuine entanglement equipartition occurs only for small charges in the relativistic case ($z=1$). For larger dynamical exponents, this equipartition is clearly violated, indicating that the presence of Lifshitz scaling strongly differentiates the charge sectors. However, the dependence of $S_1(q)$ on charge becomes progressively weaker as $z$ increases. This suggests that an effective equipartition emerges in the large-$z$ limit, where $\Delta S_1(q)$ slowly approaches zero, rendering the entropy nearly insensitive to the charge sector.

\begin{figure}[h!]
	\begin{center}
\includegraphics[scale=0.78]{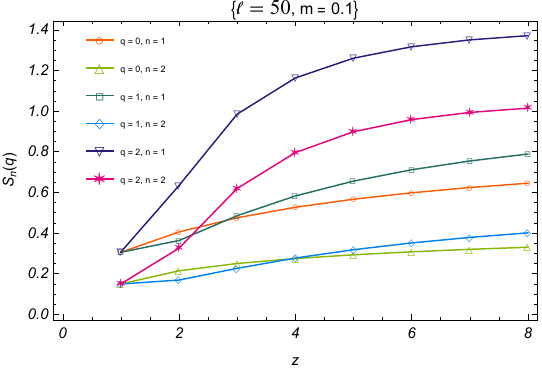}
\hspace*{1cm}
\includegraphics[scale=0.78]{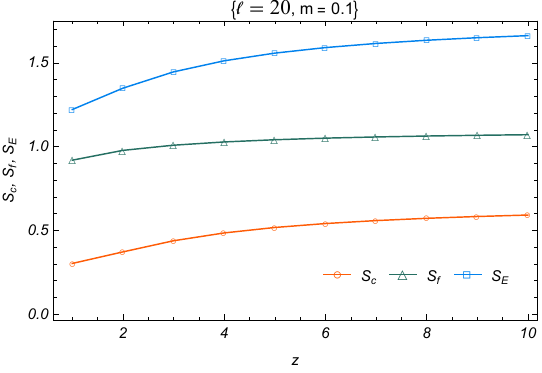}
  	\end{center}
	\caption{\textit{Left}: Symmetry resolved R\'enyi entropies as functions of the dynamical exponent for several values of the charge and R\'enyi index. \textit{Right}: Configurational, fluctuation and total entanglement entropies as function of the dynamical exponent.}
	\label{fig:SfScStotzfermionm}
\end{figure}

To conclude this section, we examine the explicit dependence of the symmetry-resolved measures on the dynamical exponent $z$, as shown in figure \ref{fig:SfScStotzfermionm}, focusing on the massive regime. The left panel presents the symmetry-resolved Rényi entropies for several values of the charge and R\'enyi index. Consistent with our earlier observations, exact equipartition occurs only in the relativistic limit ($z=1$), while for $z>1$, the symmetry-resolved entropies clearly separate across different charge sectors. This reinforces the conclusion that Lifshitz scaling generically breaks the charge equipartition property found in relativistic fermionic theories. The right panel displays the dependence of the configurational entropy and the fluctuation entropy on $z$ for various parameter choices; for completeness, the total entanglement entropy is also shown, consistent with Eq. \eqref{EEdecompose}. A striking feature, distinct from the scalar case, is that in the fermionic model $S_f$ remains larger than $S_c$ for all values of the dynamical exponent. Moreover, both quantities grow monotonically with increasing $z$ and eventually enter a saturation regime. In this limit, the inequality $S_f > S_c$ persists, indicating that the contribution of charge fluctuations within the subsystem dominates over the configurational part, while the relative weight of the operationally accessible entropy diminishes. Based on the insights of \cite{Wiseman:2003qje,Barghathi:2018idr}, this suggests that in fermionic nonrelativistic theories, the extractable or ``useful'' portion of the entanglement entropy becomes comparatively less significant as the Lifshitz exponent increases.

\subsection{Symmetry resolved entanglement in the small mass regime}

In this subsection, we compute and discuss the asymptotic expansions of symmetry-resolved measures for small values of $m$ in the Lifshitz fermionic model. As demonstrated in \cite{Vasli:2024mrf} for a massless fermionic field, whose ultraviolet (UV) fixed point respects Lifshitz scaling symmetry, the corresponding entanglement entropy in the vacuum state is independent of the dynamical exponent $z$. Here, we examine the question of how entanglement distributes across different symmetry sectors in a non-relativistic massless fermion theory. This investigation provides deeper insight into the structure of quantum correlations and the emergence of universal features in entanglement scaling. Our analysis is twofold. First, we provide a numerical analysis and examine the dependence of the measures on the dynamical exponent. Next, we will consider a toy model and show that for small subregions, the related quantities can be expressed analytically in closed form, which enables us to directly extract their scaling behavior for various parameters.

\begin{figure}[h!]
	\begin{center}
\includegraphics[scale=0.8]{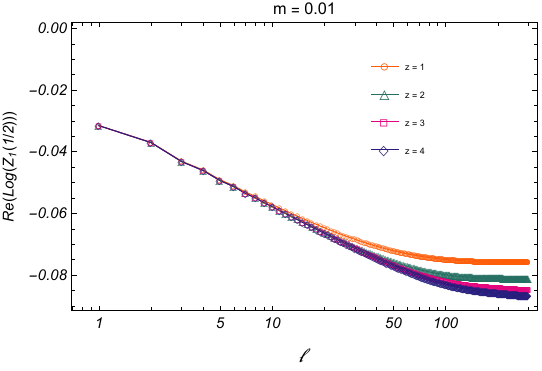}
\includegraphics[scale=0.8]{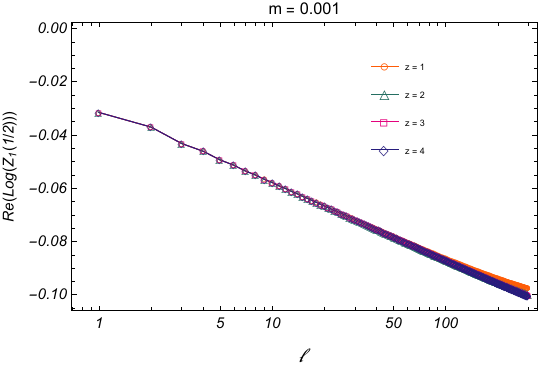}
  	\end{center}
	\caption{Real part of the logarithm of the charged moments as a function of $\ell$ for fermionic theory in the small mass limit for several values of the dynamical exponent.}
	\label{fig:Znzellfermionsmallm}
\end{figure}

\begin{figure}[h!]
	\begin{center}
\includegraphics[scale=0.8]{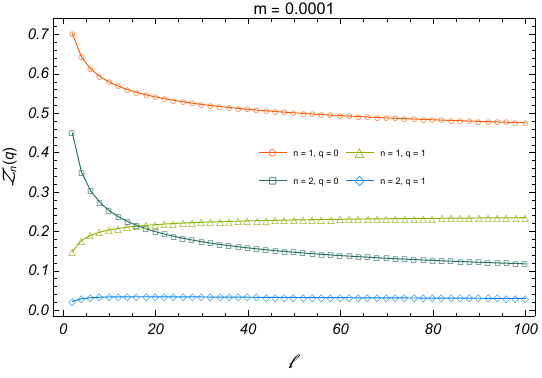}
  	\end{center}
	\caption{Symmetry-resolved partition sums as a function of $\ell$ for fermionic theory in the small mass limit. In this regime, the results are independent of the dynamical exponent and distinct curves corresponding to different values of $z$ overlap completely.}
	\label{fig:Znqzellfermionsmallm}
\end{figure}

\begin{figure}[h!]
	\begin{center}
\includegraphics[scale=0.8]{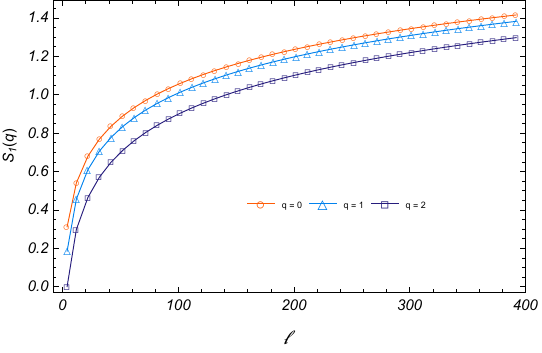}
\hspace*{1cm}
\includegraphics[scale=0.8]{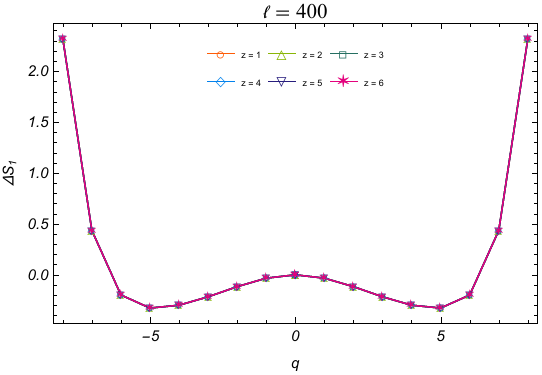}
  	\end{center}
	\caption{\textit{Left}: Symmetry-resolved entanglement entropy as a function of the entangling length for several values of the dynamical exponent and charge in the fermionic theory. \textit{Right}: Subtracted symmetry-resolved entanglement entropy as a function of charge for several values of mass and the dynamical exponent. In both panels, we fix $m\ell = 10^{-4}$.}
	\label{fig:S1qzfermionsmallm}
\end{figure}

The corresponding numerical results in the small mass limit are summarized in figures \ref{fig:Znzellfermionsmallm}, \ref{fig:Znqzellfermionsmallm} and \ref{fig:S1qzfermionsmallm}. Figure \ref{fig:Znzellfermionsmallm} illustrates the real part of the logarithm of the charged moments as a function of $\ell$ for several values of the dynamical exponent. Let us make a number of observations regarding these plots. First, we note that in both plots, the charged moment decreases as the dynamical exponent increases. Next, in the small mass limit, the dependence on $z$ becomes less pronounced, such that in the massless limit, this quantity is independent of $z$, as expected. Furthermore, in the large-subregion limit, the dependence on $\ell$ becomes logarithmic.

Figure \ref{fig:Znqzellfermionsmallm} displays the symmetry-resolved partition sums as functions of the subregion length for various values of the R\'enyi index and charge, with the mass parameter fixed at $m=10^{-4}$. Within this specific mass regime, the numerical results exhibit independence from the dynamical exponent. The numerical data reveal a fundamental qualitative distinction between the neutral and charged sectors. In the neutral sector ($q=0$), the symmetry-resolved partition sum originates from a finite value in the small-$\ell$ limit. As the subregion length increases, the partition sum decreases monotonically, eventually saturating in the large-$\ell$ limit. This behavior reflects the progressive redistribution of spectral weight among the various charge sectors as the size of the entangling region grows. In contrast, for non-zero charge sectors ($q \neq 0$), the corresponding quantities $\mathcal{Z}_n(q)$ exhibit a monotonically increasing behavior with respect to $\ell$. This contrast highlights the distinct contributions of the charged versus neutral sectors to the total entanglement structure of the system as the subregion size varies.

To further elucidate the dependence of the symmetry-resolved entanglement entropy on charge, we present the behavior of $S_1(q)$ and $\Delta S_1(q)$ across several values of the dynamical exponent and charge in Figure~\ref{fig:S1qzfermionsmallm}. While our numerical analysis reveals an apparent violation of entanglement equipartition at finite $\ell$, we find that as the entangling length $\ell$ increases (for fixed values of $m\ell$), the curves corresponding to different charge sectors slowly converge. This behavior suggests that effective equipartition emerges in the limit $m\ell \to 0$ and $\ell \to \infty$, although accessing this regime numerically is challenging due to the instability of small mass computations and the computational cost of large $\ell$.\footnote{The observed convergence is consistent with the results reported in Figure~5 of \cite{Murciano:2020vgh}, where equipartition is recovered in the scaling regime.}

To conclude our analysis, we will now investigate the dependence of the symmetry-resolved entanglement measures in the massless limit by employing an analytic treatment. As demonstrated in \cite{Vasli:2024mrf}, when the mass parameter is set to zero, the correlation function of the corresponding fermionic field, as given by Eq. \eqref{C}, simplifies to:
\begin{eqnarray}\label{Cm0}
C_{rs}=\frac{\delta_{rs}}{2}\textbf{1}+(-1)^{z+1}\frac{\sin^2\left(\frac{\pi(r-s)}{2}\right)}{i\pi(r-s)}\bm{\sigma}_3.
\end{eqnarray}
For the purpose of this simplified analysis, let us consider a toy model with a subregion length of $\ell=2$. In this specific case, the eigenvalues of the correlation matrix, denoted by $\{\nu_k\}$, become\footnote{For the sake of simplicity and clarity in the following discussion, we will focus on the scenario where $\ell=2$. However, it is important to note that the methodology can be straightforwardly extended to accommodate larger subregion lengths.}
\begin{eqnarray}\label{ell2m0nuk}
\{\nu_k\}=\{\frac{1}{2}+\frac{1}{\pi}, \frac{1}{2}-\frac{1}{\pi}\}\equiv\{\nu_+, \nu_-\}.
\end{eqnarray}
Substituting the above eigenvalues for the massless fermionic case into Eq. \eqref{Znfermion}, we obtain the following expression for the charged moments
\begin{eqnarray}\label{Znell2m0}
Z_n(\alpha)=\left(\nu_+^n+\nu_-^n\right)^2-4\left(\nu_+\nu_-\right)^n\sin^2\frac{\alpha}{2}.
\end{eqnarray}
Utilizing Eq. \eqref{znq}, it can be straightforwardly shown that, in this limit, the symmetry-resolved moments are given by
\begin{equation}\label{znqell2m0}
\mathcal{Z}_n(q)=\left(\left(\nu_+^n+\nu_-^n\right)^2-2\frac{2q^2-1}{q^2-1}\left(\nu_+\nu_-\right)^n\right)\frac{\sin \pi q}{\pi q}.
\end{equation}
Consequently, only the following charge sectors contribute to the symmetry-resolved moments:
\begin{equation}\label{znqell2m0list}
\{\mathcal{Z}_n(0), \mathcal{Z}_n(\pm 1)\}=\{\nu_+^{2n}+\nu_-^{2n}, \left(\nu_+\nu_-\right)^n\}.
\end{equation}
Interestingly, at this order of approximation, the normalization condition $\sum_q \mathcal{Z}_1(q)=1$ is exactly satisfied.\footnote{Here, we observe that $\sum_{q=-1}^{q=1} \mathcal{Z}_1(q)=\nu_+^{2}+\nu_-^{2}+2\nu_+\nu_-=(\nu_++\nu_-)^2=1$.} This finding is consistent with the interpretation of $\mathcal{Z}_1(q)$ as the probability distribution for measuring the conserved symmetry charge, yielding the value $q$. The explicit fulfillment of this normalization condition serves as a nontrivial consistency check for the zero-mass expansion. Furthermore, the numerical values derived from these expressions align with the results presented in Figure \ref{fig:Znqzellfermionsmallm}. Specifically, in the zero-mass limit, only the charge sectors with $|q|\leq 1$ exhibit non-vanishing contributions, while all other sectors vanish.
A more detailed examination reveals that, at this order, the symmetry-resolved entropy is nonzero only in the $q=0$ sector. The corresponding expression for this sector is given by
\begin{equation}\label{s1qm0}
S_1(0)=-\frac{\nu_+^2\log \nu_+^2+\nu_-^2\log \nu_-^2}{\nu_+^2+\nu_-^2}+\log\left(\nu_+^2+\nu_-^2\right).
\end{equation}
This result importantly demonstrates that, in the zero-mass limit, entanglement equipartition is absent. Consequently, the symmetry-resolved entropy, which is explicitly dependent on the charge $q$ but independent of $z$, demonstrates the absence of entanglement equipartition in this limit. This finding is consistent with the numerical results depicted in Figure \ref{fig:S1qzfermionsmallm}.

We can also extend this analysis to decompose the total entropy into configurational and fluctuation entropy in the zero mass limit. By substituting Eqs. \eqref{znqell2m0list} and \eqref{s1qm0} into Eq. \eqref{EEdecompose} we obtain
\begin{eqnarray*}\label{sfscm0}
&&S_f=-\left(\nu_+^2+\nu_-^2\right)\log\left(\nu_+^2+\nu_-^2\right)-\nu_+\nu_-\log\left(\nu_+^2\nu_-^2\right),\nonumber\\
&&S_c=\left(\nu_+^2+\nu_-^2\right)\log\left(\nu_+^2+\nu_-^2\right)-\nu_+^2\log \nu_+^2-\nu_-^2\log \nu_-^2.
\end{eqnarray*}
The key observation to note here is that the fluctuation entropy $S_f$ is the dominant contribution in this regime, while the configurational entropy $S_c$ is subleading. This behavior indicates that, in the massless case and for small $\ell$, the main contribution to the entanglement entropy comes from charge fluctuations, while the operationally accessible part, represented by $S_c$, is suppressed. Finally, the dominance of $S_f$ is consistent with the interpretation of $\mathcal{Z}_1(q)$ as the probability of measuring a given charge, reflecting that small subsystems exhibit large uncertainty in the symmetry charge and weaker structured correlations.

Further, by combining the above expressions for the configurational and fluctuation entropies, the total entanglement entropy can be written as
\begin{eqnarray*}\label{stotm0}
S_E=-\nu_-^2\log\nu_-^2-\nu_+^2\log\nu_+^2.
\end{eqnarray*}
which demonstrates that, in this limit, the total entropy is dominated by the fluctuation contribution at leading order, while the configurational part provides subleading corrections. This result smoothly interpolates between the vanishing behavior of the individual contributions and confirms the overall consistency of the expansion.

\section{Conclusions and discussions}\label{sec:conclu}

In this work, we have investigated symmetry-resolved entanglement in a variety of non-relativistic quantum field theories, including complex Lifshitz scalar chains and Lifshitz fermionic models. By employing the framework of charged moments and the correlator method, we were able to compute symmetry-resolved Rényi and von Neumann entropies and analyze their dependence on subsystem size, charge, mass, and the dynamical exponent $z$.

Our results reveal several distinctive features of non-relativistic entanglement that distinguish these systems from their relativistic counterparts. First, in Lifshitz scalar theories, we find that symmetry-resolved entanglement exhibits an approximate equipartition among charge sectors in the large-$z$ regime, which becomes more pronounced as the mass decreases. This indicates that, although Lorentz invariance is absent, the interplay of subsystem size, dynamical scaling, and charge fluctuations produces a near-uniform distribution of entanglement across sectors, effectively generalizing the equipartition property known from relativistic conformal field theories. Moreover, the decomposition of total entanglement into configurational and fluctuation contributions demonstrates that in the large-$z$ limit, the configurational entropy dominates, implying that the operationally accessible portion of entanglement becomes increasingly significant.

In contrast, the behavior of Lifshitz fermionic theories displays notable differences. Here, genuine equipartition occurs only in the relativistic limit ($z=1$), while increasing the dynamical exponent introduces a clear separation between charge sectors. Interestingly, in fermionic models, fluctuation entropy consistently exceeds configurational entropy, suggesting that in non-relativistic fermionic systems, the contribution of charge fluctuations remains dominant, and the fraction of operationally accessible entanglement is comparatively suppressed. This highlights a fundamental difference in how quantum correlations are structured in non-relativistic bosonic versus fermionic systems.

Overall, our study demonstrates that symmetry-resolved entanglement provides a powerful tool for probing the intricate structure of quantum correlations in non-relativistic systems. The results uncover a rich interplay between conserved charges, subsystem size, mass, and dynamical scaling, offering new insights into the emergence of equipartition in scalar systems and the persistence of charge fluctuations in fermionic models. These findings have direct implications for experimental platforms such as cold atom setups and mesoscopic systems, where particle-number-resolved measurements can provide access to symmetry-resolved entanglement. In conclusion, symmetry-resolved entanglement in non-relativistic field theories not only generalizes fundamental properties of relativistic quantum correlations but also unveils distinct structural features unique to Lifshitz systems, providing a comprehensive framework for studying entanglement in a broad class of quantum many-body models.

Future work may extend our analysis to interacting non-relativistic systems, higher-dimensional setups, and finite-temperature scenarios, where symmetry-resolved entanglement could reveal additional universal features and novel scaling laws. Furthermore, understanding the operational significance of configurational versus fluctuation entropy in practical quantum information protocols represents an exciting avenue for both theoretical and experimental exploration. Another interesting question is about the evolution of symmetry resolved entanglement measures after a quantum quench in Lifshitz theories. Complementarily, it would be interesting to study the notion of entanglement asymmetry and quantum Mpemba effect in such non-relativistic models which can be identified through the study of symmetry-breaking states and their relaxation under unitary dynamics \cite{Ares:2022koq,Joshi:2024sup,Rylands:2023yzx}. We will report some interesting results in these directions in early future \cite{progress}.

\subsection*{Acknowledgements}
In April 2026, we express our deep gratitude to the people of our country for their resilience and dignity in advancing and sustaining a civilization founded on the principles of humanity, rationality, and justice, and to those around the world who bear witness and give voice to this spirit. Without the support of our people, it would not have been possible to complete this work.

We would like to thank Sara Murciano and Erik Tonni for correspondence. MRMM would like to thank Mohammad Javad Vasli for related discussions. We are also very grateful to Giuseppe Di Giulio for correspondence, careful reading of the manuscript and his valuable comments. The work of M. Reza Mohammadi Mozaffar is supported by the Iran National Science Foundation (INSF) under project No. 4036941. Portions of the text were revised with the assistance of ChatGPT (OpenAI) to improve English expression.


\begin{thebibliography}{}


\bibitem{book:2000}
M. A. Nielsen, I. L. Chuang, Quantum Computation and Quantum Information, Cambridge Univ. Press., Cambridge (2000).

\bibitem{Calabrese:2009qy}
P.~Calabrese and J.~Cardy,
``Entanglement entropy and conformal field theory,''
J. Phys. A \textbf{42}, 504005 (2009)
doi:10.1088/1751-8113/42/50/504005
[arXiv:0905.4013 [cond-mat.stat-mech]].




\bibitem{Laflorencie:2015eck}
N.~Laflorencie,
``Quantum entanglement in condensed matter systems,''
Phys. Rept. \textbf{646}, 1-59 (2016)
doi:10.1016/j.physrep.2016.06.008
[arXiv:1512.03388 [cond-mat.str-el]].


\bibitem{Nishioka:2018khk}
T.~Nishioka,
``Entanglement entropy: holography and renormalization group,''
Rev. Mod. Phys. \textbf{90}, no.3, 035007 (2018)
doi:10.1103/RevModPhys.90.035007
[arXiv:1801.10352 [hep-th]].


\bibitem{Casini:2022rlv}
H.~Casini and M.~Huerta,
``Lectures on entanglement in quantum field theory,''
PoS \textbf{TASI2021}, 002 (2023)
doi:10.22323/1.403.0002
[arXiv:2201.13310 [hep-th]].

\bibitem{Rangamani:2016dms}
M.~Rangamani and T.~Takayanagi,
``Holographic Entanglement Entropy,''
Lect. Notes Phys. \textbf{931}, pp.1-246 (2017)
Springer, 2017,
doi:10.1007/978-3-319-52573-0
[arXiv:1609.01287 [hep-th]].






\bibitem{Laflorencie:2014cem}
N.~Laflorencie and S.~Rachel,
``Spin-resolved entanglement spectroscopy of critical spin chains and Luttinger liquids,''
J. Phys. A \textbf{2014}, no.11, P11013 (2014)
doi:10.1088/1742-5468/2014/11/P11013
[arXiv:1407.3779 [cond-mat.str-el]].



\bibitem{Castro-Alvaredo2024}
Olalla A. Castro-Alvaredo and Lucía Santamaría-Sanz,
\textit{Symmetry Resolved Measures in Quantum Field Theory: a Short Review},
arXiv:2403.06652 [hep-th], 2024.





\bibitem{Lukin:2019}
 A.~Lukin, M.~Rispoli, R.~Schittko, M.~E.~Tai, A.~M.~Kaufman, S.~Choi, V.~Khemani, J.~Leonard, and
M.~Greiner, ``Probing entanglement in a many-body localized system,'' Science \textbf{364}, 6437 (2019).


\bibitem{Neven:2021igr}
A.~Neven, J.~Carrasco, V.~Vitale, C.~Kokail, A.~Elben, M.~Dalmonte, P.~Calabrese, P.~Zoller, B.~Vermersch and R.~Kueng, \textit{et al.}
``Symmetry-resolved entanglement detection using partial transpose moments,''
npj Quantum Inf. \textbf{7}, 152 (2021)
doi:10.1038/s41534-021-00487-y
[arXiv:2103.07443 [quant-ph]].


\bibitem{Rath:2022qif}
A.~Rath, V.~Vitale, S.~Murciano, M.~Votto, J.~Dubail, R.~Kueng, C.~Branciard, P.~Calabrese and B.~Vermersch,
``Entanglement Barrier and its Symmetry Resolution: Theory and Experimental Observation,''
PRX Quantum \textbf{4}, no.1, 010318 (2023)
doi:10.1103/PRXQuantum.4.010318
[arXiv:2209.04393 [quant-ph]].







\bibitem{Goldstein:2017bua}
M.~Goldstein and E.~Sela,
``Symmetry-resolved entanglement in many-body systems,''
Phys. Rev. Lett. \textbf{120}, no.20, 200602 (2018)
doi:10.1103/PhysRevLett.120.200602
[arXiv:1711.09418 [cond-mat.stat-mech]].


\bibitem{Xavier:2018kqb}
J.~C.~Xavier, F.~C.~Alcaraz and G.~Sierra,
``Equipartition of the entanglement entropy,''
Phys. Rev. B \textbf{98}, no.4, 041106 (2018)
doi:10.1103/PhysRevB.98.041106
[arXiv:1804.06357 [cond-mat.stat-mech]].



\bibitem{Bonsignori:2019naz}
R.~Bonsignori, P.~Ruggiero and P.~Calabrese,
``Symmetry resolved entanglement in free fermionic systems,''
J. Phys. A \textbf{52}, no.47, 475302 (2019)
doi:10.1088/1751-8121/ab4b77
[arXiv:1907.02084 [cond-mat.stat-mech]].


\bibitem{Tan:2019axb}
M.~T.~Tan and S.~Ryu,
``Particle number fluctuations, R{\'e}nyi entropy, and symmetry-resolved entanglement entropy in a two-dimensional Fermi gas from multidimensional bosonization,''
Phys. Rev. B \textbf{101}, no.23, 235169 (2020)
doi:10.1103/PhysRevB.101.235169
[arXiv:1911.01451 [cond-mat.stat-mech]].



\bibitem{Murciano:2019wdl}
S.~Murciano, G.~Di Giulio and P.~Calabrese,
``Symmetry resolved entanglement in gapped integrable systems: a corner transfer matrix approach,''
SciPost Phys. \textbf{8}, 046 (2020)
doi:10.21468/SciPostPhys.8.3.046
[arXiv:1911.09588 [cond-mat.stat-mech]].






\bibitem{Capizzi2020}
Luca Capizzi, Paola Ruggiero, Pasquale Calabrese,
\textit{Symmetry resolved entanglement entropy of excited states in a CFT},
arXiv:2003.04670 [hep-th], 2020.


\bibitem{Murciano:2020lqq}
S.~Murciano, P.~Ruggiero and P.~Calabrese,
``Symmetry resolved entanglement in two-dimensional systems via dimensional reduction,''
J. Stat. Mech. \textbf{2008}, 083102 (2020)
doi:10.1088/1742-5468/aba1e5
[arXiv:2003.11453 [cond-mat.stat-mech]].




\bibitem{Murciano:2020vgh}
S.~Murciano, G.~Di Giulio and P.~Calabrese,
``Entanglement and symmetry resolution in two dimensional free quantum field theories,''
JHEP \textbf{08}, 073 (2020)
doi:10.1007/JHEP08(2020)073
[arXiv:2006.09069 [hep-th]].


\bibitem{Murciano:2021djk}
S.~Murciano, R.~Bonsignori and P.~Calabrese,
``Symmetry decomposition of negativity of massless free fermions,''
SciPost Phys. \textbf{10}, no.5, 111 (2021)
doi:10.21468/SciPostPhys.10.5.111
[arXiv:2102.10054 [cond-mat.stat-mech]].



\bibitem{Capizzi:2021zga}
L.~Capizzi and P.~Calabrese,
``Symmetry resolved relative entropies and distances in conformal field theory,''
JHEP \textbf{10}, 195 (2021)
doi:10.1007/JHEP10(2021)195
[arXiv:2105.08596 [hep-th]].


\bibitem{Ares:2022hdh}
F.~Ares, S.~Murciano and P.~Calabrese,
``Symmetry-resolved entanglement in a long-range free-fermion chain,''
J. Stat. Mech. \textbf{2206}, no.6, 063104 (2022)
doi:10.1088/1742-5468/ac7644
[arXiv:2202.05874 [cond-mat.stat-mech]].


\bibitem{Jones:2022tgp}
N.~G.~Jones,
``Symmetry-Resolved Entanglement Entropy in Critical Free-Fermion Chains,''
J. Statist. Phys. \textbf{188}, no.3, 28 (2022)
doi:10.1007/s10955-022-02941-3
[arXiv:2202.11728 [quant-ph]].

\bibitem{Ghasemi:2022jxg}
M.~Ghasemi,
``Universal thermal corrections to symmetry-resolved entanglement entropy and full counting statistics,''
JHEP \textbf{05}, 209 (2023)
doi:10.1007/JHEP05(2023)209
[arXiv:2203.06708 [hep-th]].


\bibitem{Capizzi2022}
Luca Capizzi, Olalla A. Castro-Alvaredo, Cecilia De Fazio, Michele Mazzoni, Lucía Santamaría-Sanz,
\textit{Symmetry Resolved Entanglement of Excited States in Quantum Field Theory I: Free Theories, Twist Fields and Qubits},
arXiv:2203.12556 [hep-th], 2022.

\bibitem{Capizzi:2022nel}
L.~Capizzi, C.~De Fazio, M.~Mazzoni, L.~Santamar{\'\i}a-Sanz and O.~A.~Castro-Alvaredo,
``Symmetry resolved entanglement of excited states in quantum field theory. Part II. Numerics, interacting theories and higher dimensions,''
JHEP \textbf{12}, 128 (2022)
doi:10.1007/JHEP12(2022)128
[arXiv:2206.12223 [hep-th]].


\bibitem{Foligno:2022ltu}
A.~Foligno, S.~Murciano and P.~Calabrese,
``Entanglement resolution of free Dirac fermions on a torus,''
JHEP \textbf{03}, 096 (2023)
doi:10.1007/JHEP03(2023)096
[arXiv:2212.07261 [hep-th]].

\bibitem{DiGiulio:2022jjd}
G.~Di Giulio, R.~Meyer, C.~Northe, H.~Scheppach and S.~Zhao,
``On the boundary conformal field theory approach to symmetry-resolved entanglement,''
SciPost Phys. Core \textbf{6}, 049 (2023)
doi:10.21468/SciPostPhysCore.6.3.049
[arXiv:2212.09767 [hep-th]].

\bibitem{Northe:2023khz}
C.~Northe,
``Entanglement Resolution with Respect to Conformal Symmetry,''
Phys. Rev. Lett. \textbf{131}, no.15, 151601 (2023)
doi:10.1103/PhysRevLett.131.151601
[arXiv:2303.07724 [hep-th]].


\bibitem{DiGiulio:2023nvz}
G.~Di Giulio and J.~Erdmenger,
``Symmetry-resolved modular correlation functions in free fermionic theories,''
JHEP \textbf{07}, 058 (2023)
doi:10.1007/JHEP07(2023)058
[arXiv:2305.02343 [hep-th]].


\bibitem{Murciano:2023ofp}
S.~Murciano, J.~Dubail and P.~Calabrese,
``More on symmetry resolved operator entanglement,''
J. Phys. A \textbf{57}, no.14, 145002 (2024)
doi:10.1088/1751-8121/ad30d1
[arXiv:2309.04032 [cond-mat.stat-mech]].


\bibitem{Caputa:2025mii}
P.~Caputa, G.~Di Giulio and T.~Q.~Loc,
``Growth of block-diagonal operators and symmetry-resolved Krylov complexity,''
Phys. Rev. Res. \textbf{7}, no.4, 4 (2025)
doi:10.1103/9v9v-54zv
[arXiv:2507.02033 [hep-th]].


\bibitem{Northe:2025qcv}
C.~Northe,
``Fermion parity resolution of entanglement,''
JHEP \textbf{12}, 134 (2025)
doi:10.1007/JHEP12(2025)134
[arXiv:2509.03605 [hep-th]].



\bibitem{Ares:2026vjt}
F.~Ares, J.~Das and A.~Kundu,
``Symmetry Resolved Entanglement Entropy: Equipartition under Driven and Non-unitary Evolution in a Compact Boson CFT,''
[arXiv:2603.28567 [hep-th]].

\bibitem{Zhao:2020qmn}
S.~Zhao, C.~Northe and R.~Meyer,
``Symmetry-resolved entanglement in AdS$_{3}$/CFT$_{2}$ coupled to U(1) Chern-Simons theory,''
JHEP \textbf{07}, 030 (2021)
doi:10.1007/JHEP07(2021)030
[arXiv:2012.11274 [hep-th]].


\bibitem{Weisenberger:2021eby}
K.~Weisenberger, S.~Zhao, C.~Northe and R.~Meyer,
``Symmetry-resolved entanglement for excited states and two entangling intervals in AdS$_{3}$/CFT$_{2}$,''
JHEP \textbf{12}, 104 (2021)
doi:10.1007/JHEP12(2021)104
[arXiv:2108.09210 [hep-th]].

\bibitem{Zhao:2022wnp}
S.~Zhao, C.~Northe, K.~Weisenberger and R.~Meyer,
``Charged moments in W$_{3}$ higher spin holography,''
JHEP \textbf{05}, 166 (2022)
doi:10.1007/JHEP05(2022)166
[arXiv:2202.11111 [hep-th]].

\bibitem{Alexandre:2011kr}
J.~Alexandre,
``Lifshitz-type Quantum Field Theories in Particle Physics,''
Int. J. Mod. Phys. A \textbf{26}, 4523-4541 (2011)
doi:10.1142/S0217751X11054656
[arXiv:1109.5629 [hep-ph]].

\bibitem{Baiguera2023}
Stefano Baiguera,
\textit{Aspects of Non-Relativistic Quantum Field Theories},
arXiv:2311.00027 [hep-th], 2023.



\bibitem{Fradkin:2006mb}
E.~Fradkin and J.~E.~Moore,
``Entanglement entropy of 2D conformal quantum critical points: Hearing the shape of a quantum drum,''
Phys. Rev. Lett. \textbf{97}, 050404 (2006).



\bibitem{Solodukhin:2009sk}
S.~N.~Solodukhin,
``Entanglement entropy in non-relativistic field theories,''
J. High Energy Phys. \textbf{04}, (2010) 101.



\bibitem{Nesterov:2010yi}
D.~Nesterov and S.~N.~Solodukhin,
``Gravitational effective action and entanglement entropy in UV modified theories with and without Lorentz symmetry,''
Nucl. Phys. \textbf{B842}, 141 (2011).


\bibitem{Keranen:2011xs}
V.~Keranen, E.~Keski-Vakkuri, and L.~Thorlacius,
``Thermalization and entanglement following a non-relativistic holographic quench,''
Phys. Rev. D \textbf{85}, 026005 (2012).




\bibitem{Zhou:2016ykv}
T.~Zhou, X.~Chen, T.~Faulkner, and E.~Fradkin,
``Entanglement entropy and mutual information of circular entangling surfaces in the 2  +  1-dimensional quantum Lifshitz model,''
J. Stat. Mech. \textbf{1609}, (2016) 093101.


\bibitem{MohammadiMozaffar:2017nri}
M.~R.~Mohammadi Mozaffar and A.~Mollabashi,
``Entanglement in Lifshitz-type quantum field theories,''
J. High Energy Phys. \textbf{07}, (2017) 120.

\bibitem{He:2017wla}
T.~He, J.~M.~Magan and S.~Vandoren,
``Entanglement entropy in Lifshitz theories,''
SciPost Phys. \textbf{3}, 034 (2017).



\bibitem{MohammadiMozaffar:2017chk}
M.~R.~Mohammadi Mozaffar and A.~Mollabashi,
``Logarithmic negativity in Lifshitz harmonic models,''
J. Stat. Mech. \textbf{1805}, (2018) 053113.

\bibitem{MohammadiMozaffar:2018vmk}
M.~R.~Mohammadi Mozaffar and A.~Mollabashi,
``Entanglement Evolution in Lifshitz-type Scalar Theories,''
JHEP \textbf{01}, 137 (2019)
doi:10.1007/JHEP01(2019)137
[arXiv:1811.11470 [hep-th]].

\bibitem{MohammadiMozaffar:2019gpn}
M.~R.~Mohammadi Mozaffar and A.~Mollabashi,
``Universal Scaling in Fast Quenches Near Lifshitz-Like Fixed Points,''
Phys. Lett. B \textbf{797}, 134906 (2019)
doi:10.1016/j.physletb.2019.134906
[arXiv:1906.07017 [hep-th]].


\bibitem{Angel-Ramelli:2020wfo}
J.~Angel-Ramelli, C.~Berthiere, V.~G.~M.~Puletti and L.~Thorlacius,
``Logarithmic negativity in quantum Lifshitz theories,''
J. High Energy Phys. \textbf{09}, (2020) 011.

\bibitem{Mollabashi:2020ifv}
A.~Mollabashi and K.~Tamaoka,
``A field theory study of entanglement wedge cross section: Odd entropy,''
J. High Energy Phys. \textbf{08}, (2020) 078.

\bibitem{Mozaffar:2021nex}
M.~R.~M.~Mozaffar and A.~Mollabashi,
``Time scaling of entanglement in integrable scale-invariant theories,''
Phys. Rev. Res. \textbf{4}, no.2, L022010 (2022)
doi:10.1103/PhysRevResearch.4.L022010
[arXiv:2106.14700 [hep-th]].

\bibitem{Boudreault:2021pgj}
C.~Boudreault, C.~Berthiere and W.~Witczak-Krempa,
``Entanglement and separability in continuum Rokhsar-Kivelson states,''
Phys. Rev. Res. \textbf{4}, 033251 (2022).



\bibitem{Mintchev:2022xqh}
M.~Mintchev, D.~Pontello, A.~Sartori and E.~Tonni,
``Entanglement entropies of an interval in the free Schr\"odinger field theory at finite density,''
J. High Energy Phys. \textbf{07}, (2022) 120.

\bibitem{Mintchev:2022yuo}
M.~Mintchev, D.~Pontello, and E.~Tonni,
``Entanglement entropies of an interval in the free Schr\"odinger field theory on the half line,''
J. High Energy Phys. \textbf{09}, (2022) 090.




\bibitem{Berthiere:2023bwn}
C.~Berthiere, B.~Chen and H.~Chen,
``Reflected entropy and Markov gap in Lifshitz theories,''
J. High Energy Phys. \textbf{09}, (2023) 160.

\bibitem{Basak:2023otu}
J.~K.~Basak, A.~Chakraborty, C.~S.~Chu, D.~Giataganas and H.~Parihar,
``Massless Lifshitz field theory for arbitrary z,''
J. High Energy Phys. \textbf{05}, (2024) 284.

\bibitem{Vasli:2024mrf}
M.~J.~Vasli, K.~Babaei Velni, M.~R.~Mohammadi Mozaffar and A.~Mollabashi,
``Entanglement in Lifshitz fermion theories,''
J. High Energy Phys. \textbf{09}, (2024) 122.



\bibitem{Khoshdooni:2025ddk}
S.~Khoshdooni, K.~Babaei Velni and M.~R.~Mohammadi Mozaffar,
``Capacity of entanglement in Lifshitz theories,''
Phys. Rev. D \textbf{112}, no.2, 026027 (2025)
doi:10.1103/7cg6-m7dn
[arXiv:2505.08297 [hep-th]].




\bibitem{Wiseman:2003qje}
H.~M.~Wiseman and J.~A.~Vaccaro,
``Entanglement of Indistinguishable Particles Shared between Two Parties,''
Phys. Rev. Lett. \textbf{91}, no.9, 097902 (2003)
doi:10.1103/PhysRevLett.91.097902
[arXiv:quant-ph/0210002 [quant-ph]].


\bibitem{Barghathi:2018idr}
H.~Barghathi, C.~M.~Herdman and A.~D.~Maestro,
``R{\'e}nyi Generalization of the Accessible Entanglement Entropy,''
Phys. Rev. Lett. \textbf{121}, no.15, 150501 (2018)
doi:10.1103/PhysRevLett.121.150501
[arXiv:1804.01114 [cond-mat.quant-gas]].




\bibitem{Pirmoradian:2023uvt}
R.~Pirmoradian and M.~R.~Tanhayi,
``Symmetry-resolved entanglement entropy for local and non-local QFTs,''
Eur. Phys. J. C \textbf{84}, no.8, 849 (2024)
doi:10.1140/epjc/s10052-024-13212-8
[arXiv:2311.00494 [hep-th]].



\bibitem{Banerjee:2024ldl}
A.~Banerjee, R.~Basu, A.~Bhattacharyya and N.~Chakrabarti,
``Symmetry resolution in non-Lorentzian field theories,''
JHEP \textbf{06}, 121 (2024)
doi:10.1007/JHEP06(2024)121
[arXiv:2404.02206 [hep-th]].

\bibitem{Gentile:2025nwv}
F.~Gentile,
``Aspects of entanglement in holography, harmonic lattices and non-relativistic quantum field theories,''
PhD thesis, SISSA, Trieste (2025),
URN: 20.500.11767/149610.

\bibitem{Peschel:2002yqj}
I.~Peschel,
``Calculation of reduced density matrices from correlation functions,''
J. Phys. A \textbf{36}, no.14, L205 (2003)
doi:10.1088/0305-4470/36/14/101
[arXiv:cond-mat/0212631 [cond-mat]].


\bibitem{Peschel:2004qbn}
I.~Peschel,
``On the entanglement entropy for an XY spin chain,''
J. Stat. Mech. \textbf{0412}, P12005 (2004)
doi:10.1088/1742-5468/2004/12/P12005
[arXiv:cond-mat/0410416 [cond-mat.stat-mech]].


\bibitem{Eisler:2009vye}
V.~Eisler and I.~Peschel,
``Reduced density matrices and entanglement entropy in free lattice models,''
J. Phys. A \textbf{42}, no.50, 504003 (2009)
doi:10.1088/1751-8113/42/50/504003
[arXiv:0906.1663 [cond-mat.stat-mech]].

\bibitem{Ares:2022koq}
F.~Ares, S.~Murciano and P.~Calabrese,
``Entanglement asymmetry as a probe of symmetry breaking,''
Nature Commun. \textbf{14}, no.1, 2036 (2023)
doi:10.1038/s41467-023-37747-8
[arXiv:2207.14693 [cond-mat.stat-mech]].

\bibitem{Joshi:2024sup}
L.~K.~Joshi, J.~Franke, A.~Rath, F.~Ares, S.~Murciano, F.~Kranzl, R.~Blatt, P.~Zoller, B.~Vermersch and P.~Calabrese, \textit{et al.}
``Observing the Quantum Mpemba Effect in Quantum Simulations,''
Phys. Rev. Lett. \textbf{133}, no.1, 010402 (2024)
doi:10.1103/PhysRevLett.133.010402
[arXiv:2401.04270 [quant-ph]].


\bibitem{Rylands:2023yzx}
C.~Rylands, K.~Klobas, F.~Ares, P.~Calabrese, S.~Murciano and B.~Bertini,
``Microscopic Origin of the Quantum Mpemba Effect in Integrable Systems,''
Phys. Rev. Lett. \textbf{133}, no.1, 010401 (2024)
doi:10.1103/PhysRevLett.133.010401
[arXiv:2310.04419 [cond-mat.stat-mech]].


\bibitem{progress}
M. R. Mohammadi Mozaffar and A. Mollabashi, work in progress.


\end{thebibliography}
\end{document}